\documentclass[12pt]{iopart}
\usepackage{iopams}
\usepackage{graphicx,float}
\usepackage[colorlinks,bookmarks=false,citecolor=blue,linkcolor=red,urlcolor=blue]{hyperref}
\usepackage{color}
\newcommand{\RR}{{L}}
\newcommand{\bc}{\begin{center}}
\newcommand{\ec}{\end{center}}
\newcommand{\be}{\begin{equation}}
\newcommand{\ee}{\end{equation}}
\newcommand{\bea}{\begin{eqnarray}}
\newcommand{\eea}{\end{eqnarray}}
\eqnobysec
\def\1.2{\frac{1}{2}}

\def\H{{\mathcal{H}}}
\begin{document}
\title{Kondo Effect in Spin Chains}
\author{Nicolas Laflorencie}
\address{Laboratoire de Physique des Solides, 
Universit\'e Paris-Sud, UMR-8502 CNRS, 91405 Orsay, France}
\address{Laboratoire de Physique Th\'eorique, IRSAMC, Universit\'e Paul Sabatier, CNRS, 31062 Toulouse, France}
\author{Erik S. S\o rensen
\footnote{Permanent address: Department of Physics and Astronomy, McMaster
University, Hamilton, ON, L8S 4M1, Canada}}
\address{Department of Physics, University of Toronto, Toronto, ON, M5S 1A7, Canada}
\author{Ian Affleck}
\address{Department of Physics \& Astronomy, University of
British Columbia, Vancouver, B.C., Canada, V6T 1Z1}

\ead{laflorencie@lps.u-psud.fr, sorensen@mcmaster.ca, iaffleck@physics.ubc.ca}

\date{\today}
\begin{abstract}
It is well-known that the free electron Kondo problem can be described by a one-dimensional (1D) 
model because only the s-wave part of the electronic wave-function is affected by
the Kondo coupling.  Moreover, since only the spin degrees of freedom are
involved in the Kondo interaction, and due to spin-charge separation in 1D, 
the universal low energy long distance physics of the Kondo model also arises
when a magnetic impurity is coupled to the end of a gap-less antiferromagnetic 
$J_1-J_2$ spin-$\1.2$ chain, where $J_1(J_2)$ is the (next-)nearest neighbor coupling.
Experimental realizations of such spin chain models are possible and 
using various analytical and numerical techniques, we present a detailed
and quantitative comparison between the usual free electron Kondo model and such
spin chain versions of the Kondo problem.
For the gap-less $J_1-J_2$ spin 
chain two cases are studied, with zero next nearest neighbor coupling, $J_2=0$, and with
a critical second neighbor coupling, $J_2=J_2^c$.
We first focus on the spin chain impurity model
with a critical second neighbor antiferromagnetic exchange  
$J_{2}^{c}\simeq 0.2412$  where a bulk marginal coupling present in
the spin chain model for $J_2<J_2^c$ {\it vanishes}. There, the usual Kondo physics is recovered 
in the spin chain model in the low energy regime
(up to negligible corrections, 
dropping as powers of inverse length or energy).
At $J_2^c$ the spin chain model is not
exactly solvable and we demonstrate the equivalence to the Kondo problem by comparing 
Density Matrix Renormalization Group 
calculations on the frustrated spin chain model with exact
Bethe Ansatz calculations of the electronic Kondo problem.
We then analyze the 
nearest-neighbor model ($J_2=0$) where
a new kind of Kondo effect occurs due to the presence of the bulk marginal coupling. 
This marginal coupling alters slightly the 
$\beta$-function for the Kondo coupling leading to a slower variation of the 
Kondo temperature $T_K$ with the bare Kondo coupling. In the exact Bethe ansatz
solution of this spin chain impurity model ($J_2=0$) Frahm and Zvyagin noted this
relation as well as the connection to the Kondo problem.  Here, by numerically solving the Bethe
ansatz equations we provide further evidence for the connection to Kondo physics
and in addition we present low temperature
Quantum Monte Carlo results for the impurity susceptibility that further support this
connection.  
\end{abstract}
\maketitle

\section{Introduction}
\begin{figure}[!ht]
\begin{center}
\includegraphics[width=10cm,clip]{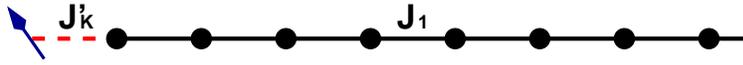}
\end{center}
\caption{Schematic picture for an open Heisenberg chain coupled to a spin impurity (arrow) at the left boundary via a 
weak antiferromagnetic exchange $J_K'$ (dashed bond). }
\label{fig:NNchain}
\end{figure}

The usual Kondo Hamiltonian~\cite{KondoRMP,Hewson} contains a Heisenberg interaction between an impurity spin 
and  otherwise non-interacting electrons. 
A simple model takes a free electron dispersion relation and a 
$\delta$
-function Kondo interaction:
\begin{equation}
H=\int d^{3}r[\psi ^{\dagger }(-\nabla ^{2}/2m)\psi +J _{K}\delta^3(\vec
r)\psi ^{\dagger }(\vec{\sigma}/2)\psi \cdot \vec{S}_{imp}].  \label{H3D}
\end{equation}
Upon expanding in harmonics, this 
can generally be reduced to a one-dimensional model of electrons 
on a semi-infinite line interacting with the impurity spin at the origin~\cite{Sorensen07}. 
We shall refer to this model as the standard free electron Kondo model (FEKM).
A closely related model consists of an open gap-less spin-$\1.2$ Heisenberg chain, 
defined on a semi-infinite line, with one coupling at the 
end of the chain weaker than the others, 
described by the following Hamiltonian 
(see Fig.~\ref{fig:NNchain}): 
\be
\H=J_1\sum_{i=1}^{L-1}{\vec{S}}_{i}\cdot{\vec{S}}_{i+1}+\H_{imp},\ \ \ 
\H_{imp}=J_K' {\vec{S}}_{imp}\cdot{\vec{S}}_1.
\label{NN}
\ee

More generally, it is of interest to add a frustrating second 
neighbor interaction, $J_2$, to this gap-less spin Hamiltonian while keeping the open boundary conditions.  
In this case we arrive at a slightly generalized version of the SCKM:
\begin{eqnarray}
&&{\cal H} =
J_{1}\sum_{i=1}^{L-1}\vec{S}_{1}\cdot \vec{S}_{i+1}+J_{2}\sum_{i=2}^{L-2}\vec{%
S}_{i}\cdot \vec{S}_{i+2} + {\cal H}_{imp},  \nonumber\\
&&{\cal H}_{imp}=J_{K}'{\vec{S}}_{imp}\cdot \left(J_1\vec{S}_{1}+J_{2}\vec{S}_{2}\right).
\label{eq:spinch}
\end{eqnarray}
\begin{figure}[!ht]
\begin{center}
\includegraphics[width=10cm,clip]{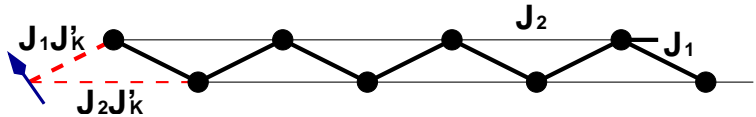}
\end{center}
\caption{Schematic picture for the open frustrated Heisenberg chain coupled to a spin impurity (arrow)
[model (\ref{eq:spinch})]. Thin lines show the frustrating second neighbor couplings $J_{2}$.}
\label{fig:NNN}
\end{figure}
In the following we shall always set $J_1\equiv 1$.
We refer to both spin chain models, Eqs.~(\ref{NN}) and (\ref{eq:spinch}), 
as the Spin Chain Kondo Model (SCKM). As $J_2$ is increased a critical point
$J_2^c\approx 0.2412$~\cite{Eggert96} is reached beyond which the spin chain enters a dimerized
phase~\cite{Haldane82} with a gap and the relation between
Eq.~(\ref{eq:spinch}) and Kondo physics no longer holds. 

The purpose 
of this paper is to explore in some detail the relationship 
between the FEKM and the SCKM. Such a relationship 
was pointed out in \cite{Eggert92,Rommer00} using field theoretical techniques 
and numerical solutions of finite chains.  In \cite{Frahm97}, hereafter 
referred to as FZ, it 
was pointed out that the spin Hamiltonian of Eq. (\ref{NN}) (or Eq.~(\ref{eq:spinch}) with $J_2=0$)
is Bethe ansatz integrable and a number of its properties were 
determined exactly, establishing some connection with the usual 
(free electron) Kondo model. However, certain clear differences 
between the two models are evident from the exact solution. 
In particular, the characteristic energy scale, the Kondo 
temperature, behaves exponentially at small Kondo coupling, 
$T_K\propto e^{-c/J_K}$ (where $c$ is a constant) in 
the FEKM but has an unusual exponential 
square root behavior \cite{Frahm97} in the SCKM at $J_2=0$, 
$T_K\propto e^{-c/\sqrt{J_K'}}$. This behavior is likely generic for all $J_2<J_2^c$. 
However, as we shall see, once we tune $J_2$ to 
the special critical point
$J_2^c\approx 0.2412$ in the SCKM the usual exponential behavior
$T_K\propto e^{-c/J_K'}$ is recovered.

Here we will explain the subtle differences in the low energy 
properties of the FEKM and SCKM using a renormalization 
group/field theory approach. The differences arise 
from a {\it marginally irrelevant} interaction in the spin chain 
model, with marginal coupling constant $g$. This is a {\it bulk} interaction,  which does 
not occur in the FEKM and is unrelated to the Kondo interaction. 
At $J_2^c$ the marginal coupling constant vanishes, $g=0$, and the usual Kondo physics
is exactly recovered in the SCKM.

In the strong coupling limit of the usual Kondo model, the impurity spin is `screened' in the
sense that it will form a non-magnetic singlet with the rest of the system. The screening
is expected to occur on a characteristic length scale, defining the so called screening cloud:
\be
\xi_K = v/T_K\propto e^{1/\lambda_K}.
\ee
Here $v$ is the characteristic velocity of the low energy excitations. Physical quantities
with zero anomalous dimension are expected to show scaling with $\xi_K/L$. Strong evidence
for the occurrence of Kondo physics in the low-energy sector
of the SCKM can therefore be established if estimates, not only of $\xi_K$ (or equivalently $T_K$), but of the 
entire scaling functions coincide in the SCKM and FEKM. We
show that this is clearly the case for the SCKM at $J_2=J_2^c$ while for the SCKM at $J_2=0$ the logarithmic
corrections arising from the marginal coupling constant $g$ leads to scaling violations and as mentioned
above, the divergence of $T_K$ and therefore $\xi_K$ acquires an unusual exponential square root behavior.
Still, the strong coupling limit remains that of a fully screened impurity for $J_2=0$ 
($g>0$).

Our conclusions will be 
confirmed in detail 
through large-scale
numerical calculations. To this end, we have made extensive use
of the {\it two} known exact Bethe ansatz solutions, for the 
FEKM~\cite{Andrei80,Wiegman80}
{\it and} for the 
SCKM at $J_2=0$~\cite{Frahm97}.
Numerical results are then obtained
by solving the two sets of Bethe ansatz equations (BAE)
for finite systems of size $L$ at $T=0$. At finite $T$, we also obtain quantum Monte Carlo (QMC) results for the SCKM at $J_2=0$. For general $J_2$, and in particular at $J_2^c$ where the relation
between the SCKM and FEKM is most direct, 
the SCKM is not exactly solvable and a sign problem appears in QMC simulations due to the frustrating term $J_{2}$. 
Fortunately, high precision results for the SCKM at $T=0$ in this regime ($J_2\neq 0$) can be obtained
using Density Matrix Renormalization Group (DMRG) methods.

These results are useful in two ways.  First of all, the SCKM
has distinct advantages over the 
free electron version for numerical simulation 
since the Hilbert space grows much more slowly with system size, $L$, 
as $2^L$ rather than the $4^L$ behavior of the free electron model.  
(On the other hand, the Numerical Renormalization Group technique 
provides a powerful alternative numerical approach for the free electron 
model.) In particular, we presented extensive DMRG results 
on entanglement entropy for the Kondo model using the 
spin chain version in \cite{Laflorencie06,Sorensen06,Sorensen07}.  
In addition, experimental realizations 
of the SCKM are possible, in particular 
in muon spin resonance experiments \cite{Chakhalian03}.
In experiments on spin chains the coupling $J_2$ usually cannot be tuned and a detailed
understanding of the relation to the FEKM as $J_2$ is varied over experimentally realistic values 
is therefore of considerable interest.

This paper is organized as follows.  In section~\ref{sec:FieldTheory}
we briefly review the field theory and renormalization 
group (RG) treatment of both free electron and spin chain 
Kondo models. We then study the interplay of the 
marginal bulk interaction and Kondo interaction 
in the SCKM, explaining the 
origin of the exponential square root form of 
the Kondo temperature using a renormalization group approach.  
In section~\ref{sec:SCKM2} we present 
results on the SCKM with a (bulk) next nearest neighbor interaction, Eq.~(\ref{eq:spinch}), tuned 
to the critical point, 
$J_2^c\approx 0.2412$, separating gap-less and dimerized phases. 
The presence of the frustrating coupling, $J_2$, allows the bulk marginal coupling, $g$
to be tuned and at the critical point, $J_2^c$,
it vanishes, $g=0$, where excellent 
agreement is obtained with the FEKM. 
A precise relationship 
between $J_K'$ in the SCKM and the Kondo coupling, $\lambda_K$, in the FEKM 
is obtained in this case which is needed for detailed comparisons~\cite{Sorensen07,Sorensen06}. This
relation is established in sub-section~\ref{sec:numparam}. In sub-section~\ref{sec:BAEDMRGNNN} we then turn to a 
discussion of the finite-size corrections to the triplet-singlet gap
in the FEKM and SCKM (at $J_2^c$) and the resulting scaling with $\xi_K/L$. 
This allows for a convenient way of extracting $\xi_K$ and we show that estimates of
$\xi_K$ and the {\it entire} scaling function coincide in the FEKM and SCKM at $J_2=J_2^c$.
We end section~\ref{sec:SCKM2} by deriving the finite-size corrections to the ground-state
energy in sub-section~\ref{sec:EgsJ2c}. As previously shown~\cite{Sorensen07,Sorensen06}, a universal term
proportional to the Kondo length scale can be identified in the ground-state energy at intermediate values of 
the Kondo coupling. This allows for an independent and absolute estimate of $\xi_K$ in the SCKM 
in excellent agreement with those obtained from the scaling of the singlet-triplet gap.
In section~\ref{sec:SCKM} we turn to a discussion of our results for the nearest neighbor 
SCKM, Eq.~(\ref{NN}), with $J_2=0$ where $g>0$ and logarithmic corrections arising from $g$ are present.
In sub-section~\ref{sec:chiimp} we first
calculate the impurity susceptibility of 
the SCKM ($J_2=0)$ at finite temperature using QMC and compare it to 
the known result for the FEKM 
obtained from the BAE.  
In sub-section~\ref{sec:BAE} we discuss the singlet-triplet gap for 
the finite size system. 
The result is compared with weak and strong Kondo coupling perturbative 
results, supplemented by weak coupling perturbation theory in $g$ for the SCKM. 
This is shown to be a convenient quantity for extracting 
the Kondo screening cloud size, $\xi_K$.  
It compares well to the expected $v_s/T_K$ where  $v_{s}$ is the characteristic velocity. 
We then discuss the
finite size corrections to the ground-state energy for the SCKM with finite $\xi_K$ and
$g>0$ in sub-section~\ref{sec:EgsJ20}.
We close section~\ref{sec:SCKM}
by discussing Bethe ansatz results of the $T=0$ magnetization \cite{Frahm97}, $M(H)$,  
in sub-section~\ref{sec:Mimp}. We observe that, in addition 
to the standard scaling function of $H/T_K$ which occurs 
in the FEKM, additional scaling violating terms appear 
which can be expanded in $g(H)$.  
Finally, in section~\ref{sec:conclusion} we summarize our results by showing $T_K(J_K')$ as obtained for the SCKM both at $J_{2}=0$ and $J_{2}^{c}$, and compare to the formula 
obtained by FZ and to that obtained from RG calculations in section~\ref{sec:FieldTheory}.

\section{The Kondo effect: A one dimensional problem\label{sec:FieldTheory}}
\subsection{Low energy theory}
 Due to the spherically 
symmetric free particle dispersion relation of the 3D model 
in Eq. (\ref{H3D}), we may
expand the electron annihilation operators in spherical harmonics. Due to
the $\delta$-function form of the Kondo interaction, only the $s$-wave
harmonic interacts with the impurity. (See, for example, 
\cite{Affleck90} for more details.) Assuming a weak Kondo coupling, we
may linearize the dispersion relation near the Fermi momentum, $k_F$,
yielding the 1D low energy effective Hamiltonian:
\begin{eqnarray}
H=\frac{v_F}{2\pi }\int_0^\infty dx\left[ i\psi_L^\dagger \frac{d}{dx}%
\psi_L-i\psi_R^\dagger \frac{d}{dx}\psi_R\right] 
+v_F\lambda_K \psi^\dagger_L(0)\frac{\vec \sigma}{2}\psi_L(0)\cdot \vec S.
\nonumber\\
\label{H1D}
\end{eqnarray}
Here the left and right moving Fermi fields are functions of $(v_Ft+x)$ and $%
(v_Ft-x)$ respectively and obey the (unconventionally normalized) equal time
anti-commutation relations:
\begin{equation}
\{\psi_{L/R}(x),\psi^\dagger_{L/R}(y)\} = 2\pi \delta (x-y),  \label{norm}
\end{equation}
and the boundary condition at the origin:
\begin{equation}
\psi_L(x=0)=-\psi_R(x=0).  \label{bc}
\end{equation}
[We have adopted the normalization conventions for the fermion fields and
the definition of $\lambda_K$ used in \cite{Affleck90}. $\lambda_K=\nu
J_K$ where $\nu$ is the density of states (per spin)  $\nu =
k_Fm/(2\pi^2)$.] 

A convenient way of analyzing the 1D model is to bosonize, since this
separates spin and charge degrees of freedom. First we take advantage of the
boundary conditions (\ref{bc}) and the fact that $\psi_{L/R}(t,x)=%
\psi_{L/R}(v_Ft\pm x)$, to write the right movers as the analytic
continuation of the left movers to the negative axis:
\begin{equation}
\psi_L(-x)\equiv -\psi_R(x),\ \ (x>0).
\end{equation}
Then the Hamiltonian, written in terms of left-movers only, becomes:
\begin{equation}
H=\frac{v_F}{2\pi }\int_{-\infty}^\infty dx i\psi_L^\dagger \frac{d}{dx}\psi_L
+v_F\lambda_K \psi^\dagger_L(0)\frac{\vec \sigma}{2}\psi_L(0)\cdot \vec S.
\label{H1DL}
\end{equation}
It is possible to write the kinetic energy in a form quadratic in charge and
spin currents:
\begin{equation}
J_L(x)\equiv \psi^{\alpha\dagger}_L(x)\psi_{L\alpha}(x),\ \ \vec J_L\equiv
\psi^{\alpha\dagger}_L\frac{\vec \sigma_\alpha^\beta}{2}\psi_{L\beta}(x).
\label{J}
\end{equation}
(Here repeated indices are summed.) The Hamiltonian becomes:
\begin{equation}
H=\frac{v_F}{2\pi }\int_{-\infty}^\infty dx\left[ \frac{1}{4}:J_LJ_L(x):+\frac{1%
}{3}\vec J_L\cdot \vec J_L\right]  
+v_F\lambda_K\vec J_L(0)\cdot \vec S.  \label{eq:FEKM}
\end{equation}
The spin and charge currents operators may be expressed in terms of spin and
charge boson fields and only the spin bosons couple to the impurity, $\vec S$%
. The kinetic energy terms just correspond to free massless (charge and
spin) boson kinetic energies.
\subsection{Renormalization group}
For a weak bare coupling, we can estimate $\xi_K$ as the length scale at
which the effective coupling becomes $O(1)$. From the weak coupling $\beta$%
-function, the variation of the effective Kondo coupling as we vary the
length scale is:
\begin{equation}
\frac{d\lambda_K}{d\ln L}=\beta (\lambda_K)=\lambda_K^2-\frac{1}{2}%
\lambda_K^3+\ldots\label{RGK}
\end{equation}
Here $\lambda_K(L)$ is the effective coupling at length scale $L$.  Alternatively, 
we may identify $L$ with $v_{F}/E$ where $v_{F}$ is the velocity and $E$ 
is a characteristic energy scale such as a temperature or magnetic field. 

We may integrate this equation from a length scale of the order a lattice
constant, $a$, up to $\xi_K$ with the effective Kondo coupling varying from
its bare value $\lambda_K^0$ to a value $c$, of $O(1)$. Assuming $%
\lambda_K^0 \ll 1$, this gives approximately:
\begin{equation}
\xi_K\approx \left[ae^{-1/c}\sqrt{c}\right]\frac{1}{\sqrt{\lambda_K^0}} \exp
[1/\lambda_K^0].  \label{xiKbeta1}
\end{equation}
Note that the first factor, in square brackets is simply a constant.
Including higher order terms in the $\beta$-function only leads to
relatively small corrections of the form:
\begin{equation}
\xi_K=\frac{\hbox{constant}}{\sqrt{\lambda_K}}\exp [1/\lambda_K]
\left[1+O(\lambda_K)\right].
\label{eq:XiKNNN}
\end{equation}
(Here we have dropped the subscript $0$ from $\lambda_K$ which refers to the
bare value of the Kondo coupling.)
We may solve for the effective coupling at scale $L$ in terms of $\xi_K$ 
at weak coupling (i.e. $L\ll \xi_K$):
\be \lambda_K(L)\approx {1\over \ln (\xi_K/L)}+{1\over 2}{\ln [\ln (\xi_K/L)]
\over \ln^2(\xi_K/L)}.\label{lambda(L)2}\ee

\subsection{Fermi liquid theory}

At low energies and long length scales, the effective Kondo coupling becomes
large, and the effective Hamiltonian flows to the strong coupling fixed
point. We may think of this fixed
point as one where the impurity spin is ``screened'', i.e. it forms a
singlet with a conduction electron. The remaining electrons behave, at low
energies and long length scales, as if they were non-interacting, except
that they obey a modified boundary condition reflecting the fact that they
cannot break up the singlet by occupying the same orbital as the screening
electron. This modified boundary condition corresponds to a $\pi /2$ phase
shift. Correspondingly in the spin chain Kondo model, the impurity spin gets
``adsorbed into the chain'' and no longer behaves like a paramagnetic spin
at low energies and long distances. The leading corrections to this low
energy long distance picture are described by lowest order perturbation
theory in the leading irrelevant operator at the strong coupling fixed
point. This is an interaction between the remaining conduction electrons,
near the screened impurity. (It doesn't involve the impurity itself since it
is screened and doesn't appear in the low energy effective Hamiltonian.)

This leading irrelevant operator is $\vec{J}_{L}\cdot
\vec{J}_{L}(0)$\cite{Affleck91a,Affleck91b}. 
From (\ref{eq:FEKM}), we see that this is proportional to the spin
part of the free electron energy density, $\mathcal{H}_{s,L}(0)$.  The energy density
has dimensions of (energy)/(length) so the corresponding coupling constant
in the effective Hamiltonian must have dimensions of length. On general
scaling grounds we expect it to be proportional to $\xi _{K}$. The precise
constant of proportionality simply corresponds to giving a precise
definition of what we mean by $\xi _{K}$. We adopt the convention:
\begin{equation}
H_{int}=-(\pi \xi _{K})\mathcal{H}_{s,L}(0).  \label{Hintcon}
\end{equation}
Here the subscripts $s$ and $L$ are a reminder that this is the spin only
part of the energy density for left movers. Note that if we start with a
system of length $L$ imposed OBC (with left and right movers), then we can
map into a system of periodic length $2L$ with left movers only. For the
purpose of doing first order perturbation theory in $H_{int}$ for quantities
like the susceptibility, specific heat or ground state energy, which are
translationally invariant in $0^{th}$ order, we may replace~\cite{Affleck90} $%
H_{int}$ by:
\begin{equation}
H_{int}\to -\frac{\pi \xi _{K}}{2L}\int_{-L}^{L}\mathcal{H}_{s,L}(x).
\end{equation}
This is equivalent to a length dependent reduction of the velocity:
\begin{equation}
v_{F}\to v_{F}\left[1-\frac{\pi \xi _{K}}{2L}\right].
\label{vshift}
\end{equation}
This then implies that the $T=0$ susceptibility, which is $L/(2\pi v_{F})$ in the
absence of the Kondo impurity becomes:
\begin{eqnarray}
\chi \to \frac{L}{2\pi v_{F}[1-\pi \xi _{K}/(2L)]}&\approx& \frac{L}{2\pi v_{F}}
+\frac{\xi _{K}}{4v_{F}}\nonumber\\ &=&\frac{L}{2\pi v}+\frac{1}{4T_{K}}.
\end{eqnarray}
Thus the zero temperature impurity susceptibility is $1/(4T_{K})$. It is this
form of the impurity susceptibility, simply related to the high temperature,
free spin behavior, $1/(4T)$, which motivates the definition of $\xi _{K}$
(and hence $T_{K}=v_{F}/\xi _{K}$) implied by (\ref{Hintcon}). We note that
this interaction $H_{int}$ is present even in the absence of an impurity,
for free fermions but then the coupling constant is of order a lattice
constant. Similarly, it is also present for the spin chain with no impurity
(i.e. $J_{K}^{\prime }=1$) with a coupling constant of order a lattice
constant. The effect of a weak Kondo coupling is to make this coupling
constant large. We emphasize that this precise choice of definition of $%
T_{K} $ has no physical consequences. The power of Fermi liquid theory is to
predict not only the form of low energy quantities but also ratios of
various low energy quantities such as impurity susceptibility, impurity
specific heat, resistivity, etc., corresponding to various generalized
Wilson ratios.

\subsection{Field theory approach to the Spin Chain Kondo Model}
\label{sec:FT}
A field theory approach to the spin chain Hamiltonian of Eq. (\ref{NN}) is 
obtained by bosonization.  The spin operators have uniform and staggered parts:
\be \vec S_j\to [\vec J_L(aj)+\vec J_R(aj)]/(2\pi )+(-1)^j\hbox{constant} \cdot \vec n(aj),\ee
where $a$ is the lattice constant and $\vec J_{L/R}$, $n$ vary slowly on that scale. 
The left and right moving current operators are equivalent to the corresponding 
operators in the free fermion model. On the other hand, the staggered spin density, $\vec n$, 
with a scaling dimension of 1/2, has no counterpart in that model. The low energy 
effective Hamiltonian is simply that of the spin part of the free electron Hamiltonian, 
\be H_0=\frac{v_s}{6\pi }\int_0^L[ \vec J_L^2+\vec J_R^2].\ee
Here $v_s$ is the spin velocity. 
In addition, there are various irrelevant operators appearing in the effective Hamiltonian. 
The only one which is important at long distances and low energies is a marginally 
irrelevant interaction,
\be H_{int}=-\frac{gv_s}{2\pi }\int_0^L\vec J_L\cdot \vec J_R.\ee
The bare coupling constant, $g_0$, has a positive value of O(1) but renormalizes 
to zero at low energies and long lengths scales. As a frustrating 
next neighbor interaction, $J_2$ is turned on $g_0$ decreases, 
passing through zero at a critical coupling, $J_2\approx 0.2412$. 
For still larger $J_2$, $g_0$ becomes negative, and hence marginally 
 relevant. 

 The RG equation for the bulk marginal coupling constant, $g$, is:
\begin{equation}
\frac{dg}{d\ln L}=\beta (g)=-g^2 -{1\over 2}g^3+\ldots 
\label{RGg}
\end{equation}
Integrating from a scale, $L_0$ of order a lattice constant 
where the bare coupling has a value $g_0>0$ to an 
arbitrary scale $L$ gives:
\be \int_{g_0}^{g(L)}{dg\over \beta (g)}=\ln (L/L_0).\label{betaint}\ee
It is known that $g(L)$ flows, at large length scales, 
to the weak coupling region where the expansion of Eq. (\ref{RGg}) 
becomes valid. For large $L$, the integral in Eq. (\ref{betaint}) 
is dominated by the small $g$ region giving:
\be {1\over g(L)}+{1\over 2}\ln g(L)=\ln (L/L_0) +\hbox{constant}\equiv \ln (L/L_1).
\label{L_1}\ee
The constant $L_1$ cannot, in general, be determined by elementary means. However, 
if the bare coupling, $g_0$ is weak then:
\be \ln (L/L_1)\approx \ln (L/L_0)+{1\over g_0},\ee
so $L_1\ll L_0$. 
For the nearest neighbor Heisenberg model, $L_1$ is O(1) and can be 
determined by comparing to the Bethe ansatz solution (for $J_K'=1$) \cite{Lukyanov}. 
In this case, the best choice of $L_1$ depends on precisely what quantity is being calculated. 
In any event, at sufficiently large $L$, where $L\gg L_1$, we may approximately 
solve Eq. (\ref{L_1}) to obtain:
\be g(L)\approx {1\over \ln (L/L_1)}-{1\over 2}{\ln [\ln (L/L_1)]\over \ln^2(L/L_1)}.
\label{g(L)2}\ee
Note that the $\beta$ functions for the Kondo coupling, $\lambda_K(L)$ 
and bulk marginal coupling, $g(L)$ are the same, to cubic order 
except for the sign of the quadratic term. Hence the 
expressions, Eq. (\ref{lambda(L)2}) and (\ref{g(L)2}) look similar. 
There are crucial differences however. Eq. (\ref{lambda(L)2})
is only valid at relatively short distances (high energies) 
where $L\ll \xi_K$; $\lambda_K(L)$ grows with increasing $L$. 
Eq. (\ref{g(L)2}) is only valid at long distances (low energies), $L\gg L_0$ 
for the nearest neighbor Heisenberg model; $g(L)$ decreases
with increasing $L$. 

  For $g_0<0$,
there is a gap of order $\sqrt{|g_0|}e^{-1/|g_0|}$, corresponding to the 
spontaneously dimerized phase. The critical point, where $g_0=0$ is 
thus an especially good point for numerical studies since, for 
$g_0>0$ the slowly decreasing marginal coupling constant complicates  finite size scaling. 

Weakly coupling an impurity spin to a single spin at a generic point in a spin chain yields 
a model \cite{Eggert92,Rommer00} with essentially no relation to the standard Kondo model. 
On the other hand, the model of Eq. (\ref{NN}) with the impurity 
spin coupled at the end of the chain, is quite different. To see this, 
first consider the limit $J_K'=g=0$.  The open boundary condition leads to:
\be \vec J_R(0)=\vec J_L(0)\propto \vec n(0).\ee
Thus the SCKM is equivalent to the FEKM for some non-trivial 
constant of proportionality between Kondo couplings, and with the Fermi 
velocity replaced by the spin velocity, ignoring 
the irrelevant operator. 

The low energy effective Hamitonian of the SCKM Eq.~(\ref{NN}) has three main terms:
\begin{eqnarray} H=\frac{v_s}{2\pi }\int_{-L}^{L}dx\frac{1}{3}\left[{\vec{J}}_L(x)\right]^2
-\frac{gv_s}{2\pi }\int_0^{L} dx \vec J_L(x)\cdot\vec J_R(x)
  +\lambda_K v_s\vec J_L(0)\cdot \vec S.\nonumber\\
\end{eqnarray}
The first term, which we denote $H_0$, corresponds to a free boson Hamiltonian, 
the second term is the marginal bulk interaction, and the third term stands for
 the Kondo like interaction at the boundary with $\lambda_K\propto J_K'$. 
Using the boundary condition, $\vec J_L(t,x=0)=\vec J_R(t,x=0)$ 
we conclude that we can regard $\vec J_R(x)$ as the analytic continuation of 
$\vec J_L(x)$ to the 
negative $x$-axis: 
\begin{equation}
\vec J_R(x)=\vec J_L(-x).\end{equation}
Hence we may write the Hamiltonian as:
\begin{equation} H=H_0
-\frac{gv_s}{2\pi }\int_{0}^{L} dx \vec J_L(x)\cdot\vec J_L(-x)
+\lambda_K v_s\vec J_L(0)\cdot \vec S.
\end{equation}
When $g=0$ this is the same field theory that describes the FEKM, Eq.~(\ref{eq:FEKM}), with $v_F$ replaced by $v_s$, the 
spin velocity. 
A non-zero $g$ leads to some differences, but since $g$ is 
marginally irrelevant, its effects becomes progressively less important 
at long distances and low energies. 

We now consider the renormalization group equations for the model 
with both $g$ and $\lambda_K\neq 0$. 
The presence of the Kondo interaction, which is only at the boundary,
 cannot change the renormalization of the 
bulk interaction, $g$. On the other hand, the reverse is not true; the 
bulk marginal interaction has an important effect on the renormalization 
of the Kondo coupling constant. This can be viewed as resulting from 
the fact that the boundary operator $\vec J_L(0)$, appearing in the 
Kondo interaction, which has a scaling dimension of $1$ when $g=0$ picks 
up an anomalous dimension of first order in $g$.  This anomalous 
dimension was calculated in \cite{AQ}:
\be x= 1-g+\ldots \ee
[Note that, in \cite{AQ}, $g$ was normalized differently, so 
that $g_{AQ}=(\sqrt{3}/4\pi )g$.]
This observation determines the RG equations:
\begin{equation} \frac {d\lambda_K}{d\ln L}=g\lambda_K+\lambda_K^2 + \ldots \label{betag}\end{equation}
To solve (\ref{betag}) we may substitute $g(L)$ from 
(\ref{g(L)2}), yielding:
\begin{equation}
\frac{d\lambda_K}{d\ln L}=\frac{\lambda_K}{\ln (L/L_1)}+\lambda_K^2.\end{equation}
This differential equation can be solved exactly by defining a new effective 
coupling constant:
\begin{equation} \tilde \lambda (L)\equiv \frac{\lambda_K (L)}{\ln (L/L_1)},\label{ltilde}\end{equation}
which obeys the RG equation:
\begin{equation}
\frac{d\tilde \lambda}{d\ln L}=\tilde \lambda^2\ln (L/L_1).\end{equation}
 Setting 
$\lambda_{K}(\xi_{K})=1$, and using Eq.~(\ref{g(L)2}), one can obtain the renormalized expression for the effective Kondo coupling
\be
\lambda_{K}(L)\simeq \frac{\ln (L/L_{1})}{(1/2)\ln^{2}(\xi_{K}/L_{1})-(1/2)\ln^{2}(L/L_{1})+\ln(\xi_{K}/L_{1})}.
\ee
This gives, for the Kondo length scale $\xi_{K}$, the unusual dependence on the bare Kondo coupling $\lambda_{K}^{0}$:
\be
\xi_K=L_1\exp\left[-c +\sqrt{2\ln (L_0/L_1)/\lambda_{K}^{0}+
\ln^2(L_0/L_1)+c^{2}}\right],
\label{xiK}\end{equation}
where $c$ is a positive constant of O(1).

Clearly, $\lambda_{K}(L)$ is not a scaling function of $\xi_{K}/L$. Nevertheless, when the bare marginal coupling $g_{0}$ is very small, 
which is true when $J_2$ is tuned close to the critical point, the length scale 
\be
L_{1}\sim \exp(-1/g_{0})\ll L_0
\label{L1}
\ee
and then 
\be
\lambda_{K}\to \frac{1}{\ln (\xi_{K}/L)}.
\ee

Assuming $\lambda_{K}^{0}\ll 1/\ln (L_0/L_1)$, with a fixed bare marginal coupling $g_{0}$ Eq.~(\ref{xiK}) may be approximated:
\begin{equation}
\xi_K \approx L_1\exp \sqrt{2\ln (L_0/L_1)/\lambda_{K}^{0}}.\label{xiKapprox}
\end{equation}
  Note 
that this dependence of $\xi_K$ on the Kondo coupling, 
$\xi_K\propto \exp [b/\sqrt{\lambda_{K}^{0}}]$ (where $b$ is a constant), gives a much shorter 
Kondo length scale than in the usual Kondo effect  which occurs when $g_{0}=0$:
$\xi_K\propto \exp [1/\lambda_{K}^{0}]$.
Also note, from Eq.~(\ref{L1}) that the limit of zero marginal coupling 
corresponds to $L_1\to 0$. In this limit, $\ln (a/L_1)\gg 1/\lambda_{K}^{0}$ 
so we may Taylor expand the square root in (\ref{xiK}) giving 
$\xi_K\sim \exp [1/\lambda_{K}^{0}]$. However, 
(\ref{xiK}) and (\ref{xiKapprox}) 
only hold when $\xi_K\gg L_1$ where $L_1$ is roughly the scale 
at which the effective marginal bulk coupling, $g(L)$ starts to become small. 

We expect physical quantities with zero anomalous dimension to be generalized scaling 
functions.  This means that they can be written as functions of 
$\xi_K/L$ and $g(L)$ only. In general, they should have a Taylor expansion 
in the renormalized coupling $g(L)$:
\be f(\lambda_K,g,L)=\sum_{n=0}^\infty g^n(L)f_n(\xi_K/L).\label{scalcorr}\ee
The leading term, $f_0(\xi_K/L)$ should be the same as in the model 
with $g=0$, the FEKM.  
Note that, due to the presence of the $g^n(L)\sim 1/\ln^n(L)$, $f$ is not a pure scaling 
function in the sense that it does not depend {\it only} on $\xi_K/L$, but 
since $g(L)\to 0$ as $L\to \infty$, $\xi_K/L$ scaling 
becomes a better and better approximation at large $L$. 
More generally, the inverse length $1/L$ in Eq. (\ref{scalcorr}) may be 
replaced by an energy scale such as a magnetic field.

The modified (exponential square root) expression for the 
dependence of the Kondo temperature on Kondo coupling 
and the slowly decreasing correction, $gf_1$, $g^2f_2,\ldots$ are 
the two main differences between the FEKM and the SCKM. 
In the following sections we demonstrate the correctness 
of this assertion by considering various physical quantities. 

\section{Kondo effect in the SCKM at $J_2^c$\label{sec:SCKM2}}

We first discuss the frustrated spin chain model (depicted in Fig.~\ref{fig:NNN}) with the Hamiltonian 
Eq~(\ref{eq:spinch}). 
As we have already stressed, in this case the marginal coupling is {\it exactly zero}, $g=0$,
and the mapping to the FEKM becomes exact, up to strictly irrelevant interactions. 
Hence, in this case, by determining various constants numerically, it is possible to establish a precise
mapping between the low-energy, long-distance sectors of the FEKM and SCKM. 
In sub-section~\ref{sec:numparam}
we numerically determine  $v_s$ in the SCKM as well as the constant of proportionality
between $J_K'$ in the SCKM and $\lambda_K=\nu J_K$ in the FEKM. Then, in sub-section~\ref{sec:BAEDMRGNNN}, we
directly
compare Bethe ansatz results for the FEKM with DMRG results for the SCKM at $J_2^c$. Finally, in sub-section~\ref{sec:EgsJ2c} we
show how $\xi_K$ at intermediate couplings, $J_K'$, for the SCKM can be determined from the finite-size scaling of the ground-state energy in the SCKM.

First consider the model with $J_K'=0$. 
 At a general site, $j$, the low
energy degrees of freedom of the spin operators are represented as:
\begin{equation}
\vec S_j\approx \frac{1}{2\pi}[\vec J_L(aj)+\vec J_R(aj)]+(-1)^j%
\hbox{constant}\times \vec n(aj).
\end{equation}
Here $a$ is the lattice spacing and $\vec n(aj)$, the alternating part of
the spin operators can be written in terms of the spin boson field in a
non-linear way. At the end of the chain, due to the open boundary condition,
we find that $\vec n(x)\to \hbox{constant}\times \vec J_L(0)$ and therefore:
\begin{equation}
\vec S_2+J_2\vec S_3 \approx C\vec J_L(0),  \label{Cdef0}
\end{equation}
where $C$ is a non-universal constant, depending on the second neighbor
coupling, $J_2$ in the Hamiltonian. ($C$ has dimensions of inverse length,
and is proportional to the inverse lattice spacing.) Now including a weak
coupling, $J_K'$ to the first spin, the low energy effective
Hamiltonian becomes the usual FEKM with the replacements:
\begin{equation}
v_F \to v_s,\ \ v_F\lambda_K \to CJ_K'.  \label{lambdaJ}
\end{equation}
\subsection{Numerical parameters for the spin chain problem}\label{sec:numparam}
{\it Velocity of excitations:}
In order to make the above correspondence quantitative, one needs
to determine exactly the spin velocity $v_s$ at the critical point $J_2=J_{2}^{c}$.
We can use the fact that the finite size scaling of the ground state energy for the uniform SCKM with $J_K'=1$ is simply given by~\cite{Blote86}
\be
E_0(L)=\epsilon_0 L+\epsilon_1-\frac{\pi c v_s}{24L}+\frac{a_2}{L^2}+\frac{a_3}{L^3}+O(1/L^4).
\label{eq:E_OBC}
\ee
$\epsilon_0,~\epsilon_1$ and $a_2,~a_3$ are non-universal numbers, whereas the $1/L$ term is universal, proportional to the central charge ($c=1$ here)
and the spin velocity $v_s$. Note that, $v_s$ will depend on $J_2$ and even though it is known analytically from the Bethe ansatz solution available
at $J_2=0$ we need to determine it at $J_2^c$ where no exact results are available. 
\begin{figure}[!ht]
\begin{center}
\includegraphics[width=10cm,clip]{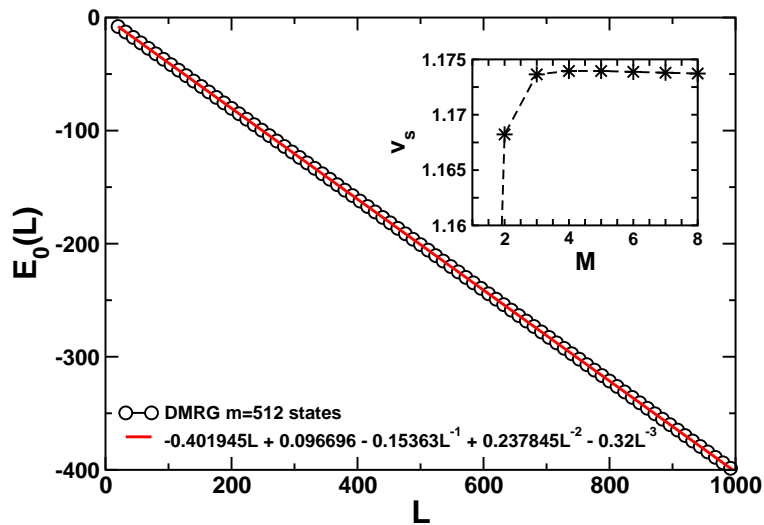}
\end{center}
\caption{Finite size scaling of the ground state energy $E_0(L)$ of the 
	SCKM at the critical second neighbor coupling $J_{2}^{c}=0.2412$ shown 
	for chains up to $L=1000$ sites.	
	DMRG results obtained keeping $m=512$ states (open symbols) are compared to the finite
	size scaling form~(\ref{eq:E_OBC}) up to third order (full line). 
	In the inset, $v_s$ obtained from polynomial fits of the form 
	$E_0(L)=\epsilon_0 L + \epsilon_1 -\pi v_s/(24L) + \sum_{i=2}^{M}a_i/L^i$ 
	is plotted versus $M$.}
	
\label{fig:E_OBC}
\end{figure}
DMRG results for $E_0(L)$, computed 
on chains up to $L=1000$ sites, are shown in Fig.~\ref{fig:E_OBC} 
and fitted to Eq. (\ref{eq:E_OBC}). We have checked that fitting with a $M^{\rm{th}}$-order
polynomial form in $1/L$ does not change the estimate of $v_s$ whenever $M\ge 3$. Thus we found for 
the spin velocity
\be
v_s=1.174(1).
\ee
This is in good agreement with previous estimates~\cite{Eggert92}.

{\it Kondo coupling:}
Of course, to predict $\xi_K$ for the spin chain model we need to
exploit Eq. (\ref{lambdaJ}) to
relate 
$\lambda_K$ to $J_K'$, the weak coupling of the spin at the end of the
chain, where the constant $C$ is determined by (\ref{Cdef0}).
We emphasize that this constant $C$ can be determined in the
theory with zero Kondo coupling, $J_K'=0$ and we can therefore study correlations in chains
of even length. A convenient way of
determining $C$ is to measure a long-distance correlation function in the
spin chain and compare it to the corresponding correlation function in the
continuum field theory. The simplest choice appears to be the end-to-end
equal time correlation function:
\begin{equation}
\langle(\vec S_2+J_2\vec S_3)\cdot (\vec S_{L}+J_2\vec S_{L-1})\rangle \approx  C^2\langle\vec J_L(0)\cdot \vec J_L(L)\rangle.  \label{Cdef}
\end{equation}
The current correlation function is simply that of the free fermion model
with the current defined in Eq. (\ref{J}) and the fermion fields normalized
as in Eq. (\ref{norm}). For an infinite chain this implies:
\begin{equation}
\langle\vec J_L(r)\cdot \vec J_L(r^{\prime})\rangle=-\frac{3}{2(r-r^{\prime})^2}.
\end{equation}
However, we need the correlation function for a finite strip. Using the fact
that the open boundary conditions on a finite strip of length $L$
are equivalent to periodic boundary conditions for left movers only on a
circle of circumference $2L$, we obtain:
\begin{equation}
\langle\vec J_L(r)\cdot \vec J_L(r^{\prime})\rangle=-\frac{3}{(8L^2/\pi^2
)\sin^2 [\pi (r-r^{\prime})/2L]},
\end{equation}
giving the end-to-end correlation function:
\begin{equation}
\langle\vec J_L(0)\cdot \vec J_LL)\rangle=-\frac{3\pi^2}{8L^2}.
\label{e2eft}
\end{equation}
\begin{figure}[!ht]
\begin{center}
\includegraphics[width=10cm,clip]{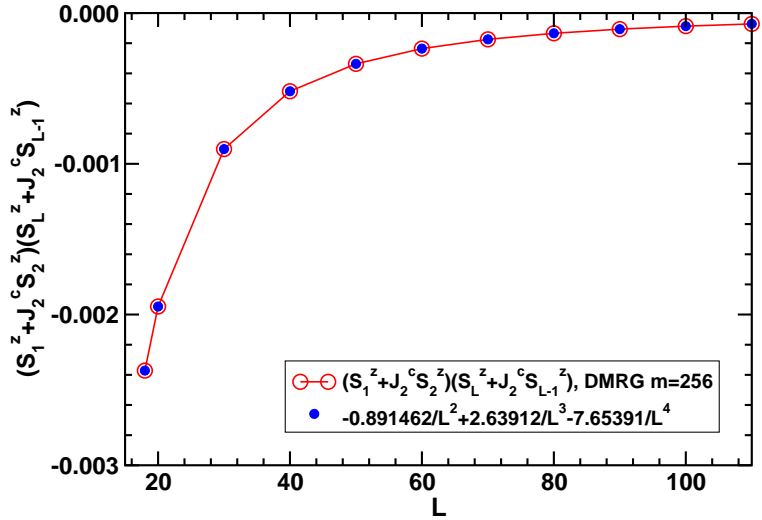}
\end{center}
\caption{DMRG results keeping $m=256$ states for $\langle(\vec S_2^z+J_2\vec S_3^z)\cdot (\vec S_{%
L}^z+J_2\vec S_{L-1}^z)\rangle$ for the SCKM
at $J_2^c$ (open circles). Results are shown for $L=18,20,30\ldots 110$. The solid circles indicates a polynomial fit with
the leading term of the form $-0.891462L^{-2}$. }
\label{fig:EE}
\end{figure}
As shown in Fig.~\ref{fig:EE} this correlation function can be determined
very accurately using DMRG methods, from which we determine:
\begin{equation}
\langle(\vec S_2+J_2\vec S_3)\cdot (\vec S_{L}+J_2\vec S_{L%
-1})\rangle \to -\frac{3\times 0.891462}{L^2}.  \label{e2edmrg}
\end{equation}
From Eqs. (\ref{Cdef}), (\ref{e2eft}) and (\ref{e2edmrg}) we obtain:
\begin{equation}
0.891462=\frac{\pi^2C^2}{8}, \ \ C=0.850054
\end{equation}
From (\ref{lambdaJ}), using $v_s\approx 1.174(1)$, we finally obtain the
proportionality constant between Kondo couplings in spin chain and fermion
models, $J_K'$ and $\lambda_K$:
\begin{equation}
J_K'= (v_s/C)\lambda_K=1.38~\lambda_K.
\end{equation}
\subsection{The triplet-singlet gap and Bethe ansatz results for the FEKM versus DMRG
results on the SCKM at $J_2^c$}\label{sec:BAEDMRGNNN}
\begin{figure}[t]
\begin{center}
\includegraphics[width=12cm,clip]{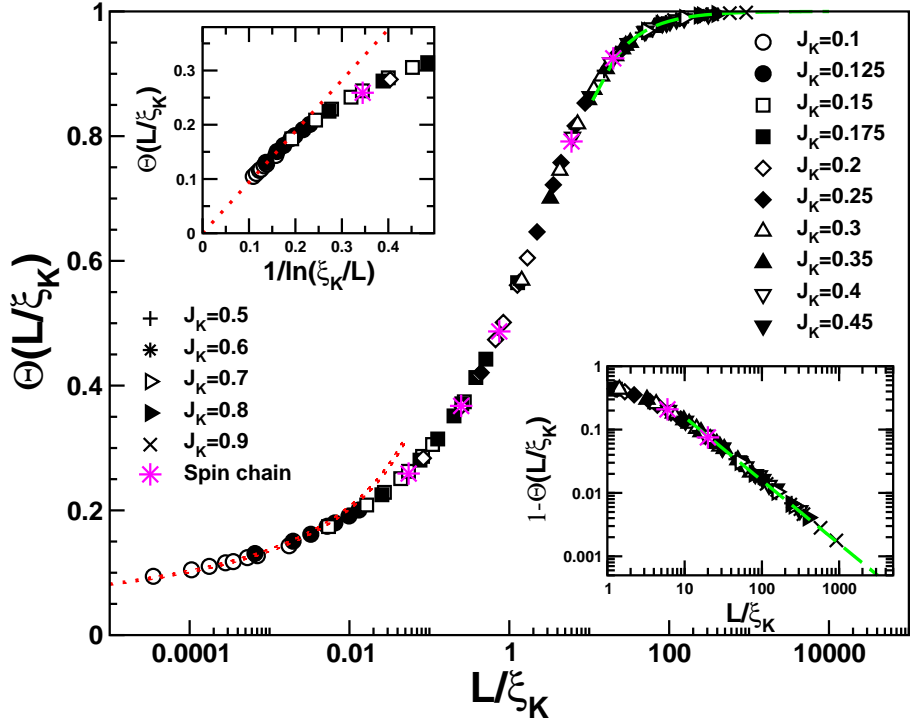}
\end{center}
\caption{
Scaling function $\Theta(L/\xi_{K})$ (see Eq.~(\ref{eq:THETA})) obtained from
the Bethe Ansatz solution of the one-channel FEKM
Hamiltonian~\cite{KondoRMP} with various values for the Kondo interaction
strength $J_K$, as indicated on the plot. The scaling plot is constructed using the strong coupling limit curve [long-dashed green line Eq.~(\ref{eq:THETA1})] as a support for the rest of the collapse.
The spin chain model
(\ref{eq:spinch}) at $J_{2}^{c}$ results (star) are adjusted to collapse onto
the universal curve. Left inset: in the weak coupling regime, the ratio
$\Theta\sim 1/\ln(\xi_{K}/L)$, as show by the linear fitting dotted line.
Right inset: in the strong coupling regime, the FLT result
$\Theta=1-\frac{\pi\xi_{K}}{2L}$ describes perfectly the behavior (long
dashed line).}
\label{fig:THETANNN}
\end{figure}

A convenient diagnostic for testing the correspondence between the FEKM and the SCKM at $J^c_2$ 
is the finite size behavior of the gap between the singlet ground state and 
lowest triplet exited state (for even $L$). Let us first consider the models with 
no Kondo coupling. Thus we consider either a free fermion chain or a Heisenberg spin 
chain with open boundary conditions.  Noting that the allowed wave-vectors are $k\approx \pi n/L$, 
 we see that the excitation energy of a triplet particle-hole excitation in the free fermion chain is:
\be \Delta_{ST}={\pi v_s\over L}.\label{eq:THETA0}\ee
The same result holds for the spin chain with open boundary conditions \cite{Eggert92}.
The extension of this result for $\xi_K\ll L$, is easily obtained from Fermi liquid 
theory, by the usual replacement Eq. (\ref{vshift}):
 \be \Delta_{ST}\to {\pi v_s\over L}\left[1-{\pi \xi_K\over 2L}\right] .\label{eq:THETA1}\ee
In the weak coupling limit we may calculate $\Delta_{ST}$ in perturbation theory 
in the Kondo coupling, $\lambda_K$. We consider a spin chain with an even number of 
sites, or equivalently a free fermion chain with an odd number of fermions, coupled 
to the impurity spin so that the ground state is a spin singlet. In the limit 
of zero Kondo coupling the ground state of the non-interacting chain is a spin doublet. 
First order perturbation theory in $\lambda_K$ couples this doublet to 
the impurity spin with a coupling constant $\pi v_s/L$, giving a singlet-triplet 
splitting of $\Delta_{ST}\approx (\pi v_s/L)\lambda_K$. We expect that, as usual,
 higher order corrections will replace $\lambda_K$ by its renormalized value 
at scale $L$, Eq. (\ref{lambda(L)2}), giving:
\be \Delta_{ST}\to  {\pi v_s\over L}{1\over \ln (\xi_K/L)}.\label{eq:THETA2}\ee
Hence, in general we expect the quantity
\be \Theta\equiv {L\over \pi v_s}\Delta_{ST},\label{eq:THETA}\ee
to be a universal scaling function of $L/\xi_K$. The same function should 
occur for the FEKM and for the SCKM up to higher order corrections 
from irrelevant operators. 

Introduced 25 years ago, independently by Andrei~\cite{Andrei80} and Wiegman~\cite{Wiegman80}, 
the exact diagonalization of the Kondo Hamiltonian (\ref{H1DL}) based on the Bethe Ansatz
provides a powerful method to access  several physical observables like 
the uniform susceptibility, the field induced magnetization, 
the specific heat, the thermodynamic entropy. 
Similarly to what was previously done using the Bethe Ansatz solution of the unfrustrated chain, one can add an external magnetic field along the $z$-axis ${\vec{B}}=B{\vec{e}}_z$, which couples to the total spin operator $S^{z}_{tot}$. Then one can solve numerically for finite length chains the Bethe Ansatz equations~\cite{Andrei80, Wiegman80} in any $S^{z}_{tot}$ sector and compute the singlet triplet excitation gap.

We have calculated $\Theta (L/\xi_K)$ for the FEKM by 
 solving the set of coupled Bethe Ansatz equations for various values of the Kondo coupling $J$, where the convention of Ref.~\cite{KondoRMP} was followed: $J=\frac{\pi}{2}\lambda_{K}$. In Fig.~\ref{fig:THETANNN} are shown the results for the ratio $\Theta (L/\xi_{K})$,
 successfully compared with the strong and weak coupling predictions Eqs.~(\ref{eq:THETA1}) and (\ref{eq:THETA2}).

We have also compared these results to DMRG calculations performed on the Kondo
spin chain model with $J_2=J_{2}^{c}$, (\ref{eq:spinch}) for
$J_{K}'=0.2,~0.25,~0.3,~0.45,~0.6$. The ratio $\Theta$ is determined for
$L=100$ sites and the DMRG data points (shown as $\star$) have been superimposed on the Bethe
Ansatz results on Fig.~\ref{fig:THETANNN}, adapting $L/\xi_{K}(J_{K}')$ in
order to get a good collapse. The resulting values for the Kondo length scales
are listed in table~\ref{tab:NNN}.

\begin{table}
\centering
\begin{tabular}{|r||ccccc|}
\hline\hline
$J_K'$&
0.6   &
0.45  &
0.3   &
0.25  &
0.2   \\
\hline
$\xi_K$
&5
&17
&130
&400
&1800\\
\hline\hline
\end{tabular}
\caption{Kondo length scale estimates for the SCKM at $J_2^c$, Eq.~(\ref{eq:spinch}).}
\label{tab:NNN}
\end{table}
%

The fact that the numerical estimates of not only the length scale $\xi_K$, but of
the {\it entire} scaling function coincide in the FEKM and the SCKM at $J_2=J_2^c$ is very
strong independent evidence for the occurrence of Kondo physics in the low-energy sector of
the SCKM at $J_2=J_2^c$.

\subsection{Finite size scaling of the ground state energy in the SCKM at $J_2^c$}\label{sec:EgsJ2c}
One way of estimating $\xi_K$, in the case $\xi_K\ll L$, is from a
universal correction to the ground state energy. This may be calculated from
the Fermi liquid theory Hamiltonian of (\ref{Hintcon}). It is well known
that the ground state energy of the open spin chain has the form:
\begin{equation}
E_{GS}=e_0+e_1L-\frac{\pi v_{s}}{24L} + \ldots
\end{equation}
where $e_0$ and $e_1$ are non-universal but the third term is universal,
depending only on $v_s$ and $\ldots$ indicates terms that drop off faster with
$L$. We may include an additional boundary term from first order
perturbation in the FLT interaction of (\ref{Hintcon}) by replacing $v_s$ by
its shifted value in (\ref{vshift}), giving:
\begin{equation}
\delta E_{GS} = \pi^2v_{s}\xi_K/(48L^2).
\end{equation}
We emphasize that such a $1/L^2$ term is always present, even for
the uniform chain with open boundary conditions ($J_K'=1$) but that the
factor of $\xi_K$ is of order the lattice spacing in that case. This term
obtains a large coefficient at weak Kondo coupling. 
Combining these contributions we find for the ground-state energy:
\begin{equation}
E_{\rm GS}(L,J_K^\prime)=e_{0}(J_K')+e_{1}L-\frac{\pi v_{s}}{24L}+\left( e_{2}+\frac{\pi
    ^{2}v_{s}\xi _{K}}{48}\right) \frac{1}{L^{2}}.
\label{eq:Egs}
\end{equation}
Please note that, while the constants $e_1,e_2$ describe bulk behavior and hence are independent
of $J_K'$, the surface term $e_0$ {\it does} depend on $J_K'$. Secondly, the difference between $e_2$ and $\xi_K(J_K'=1)$
is largely a matter of convention. Thirdly, this expression for the ground-state
energy neglects logarithmic corrections arising from the bulk marginal operator, $g$, assumed to
be zero since $J_2=J_2^c$. The more general case with $J_2<J_2^c, g>0$ will be considered in section~\ref{sec:EgsJ20}.

Using DMRG results for the ground-state energy we now attempt to determine $\xi_K(J_K')$ using
Eq.~(\ref{eq:Egs}). As outlined in \ref{app:egs} we can in this
case eliminate the 3 first terms in Eq.~(\ref{eq:Egs}) by using
a form of Richardson extrapolation. We then arrive at:
\begin{eqnarray}
E^{(3)}(L,J'_K)&=&\left( e_{2}+\frac{\pi ^{2}v_{s}\xi _{K}}{48}\right)c^3_2(L)+\mathcal{O}(L^{-1})\nonumber\\
&=&Dc^3_2(L)+\mathcal{O}(L^{-1})
\end{eqnarray}
where the coefficient $c^3_2(L)$ is known. Plotting $E^{(3)}/c^3_2$ should then yield the desired constant, $D$.
\begin{figure}[!ht]
\begin{center}
\includegraphics[width=10cm,clip]{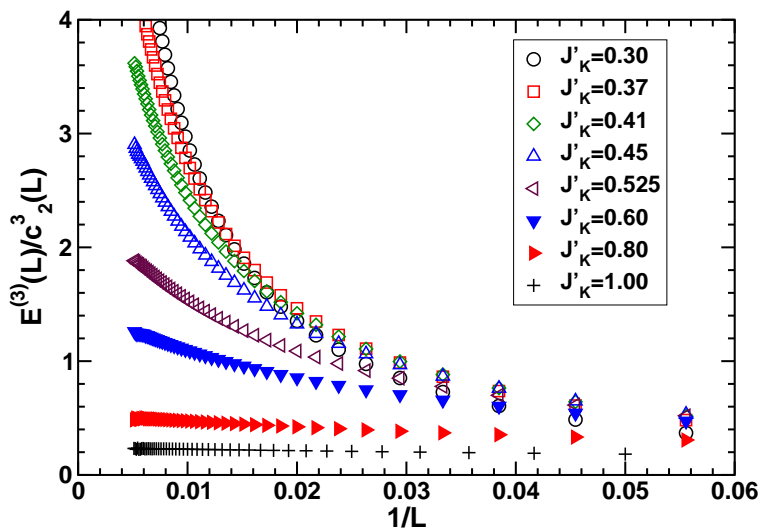}
\end{center}
\caption{DMRG results keeping $m=256$ states ($m=512$ for $J_K'=1$) for $E^{(3)}(L,J'_K)/c^3_2(L)$
  for the SCKM at $J_2^c$ for a range of $J'_K$ plotted versus $1/L$.
For the larger $J'_K$ the results quickly approach a constant as $1/L\to 0$.
Results are obtained from the ground-state energy for systems with even size.
  }
\label{fig:GSErg}
\end{figure}
Plots of $E^{(3)}/c^3_2(L)$ versus
$1/L$ are shown in Fig.~\ref{fig:GSErg}. For $J'_K=0.6,0.8,1.0$ it is easy to
extract the constant $D$ using a simple polynomial fit. However, for smaller $J'_K$ we have found
it necessary to fit to a (2,2) rational polynomial to improve the extrapolation. For $J_K'$ smaller than
$J'_K=0.3$ it is no longer possible
to extrapolate using rational polynomials and a reliable numerical determination of $D$ becomes impossible.  

As mentioned above, we expect $\xi_K$ to be $\mathcal{O}(1)$ for $J'_K=1$. 
In previous work~\cite{Sorensen07,Sorensen06} we used Fermi liquid theory to determine $\xi_K(J_K'=1)=0.65$
and here we use this result to calibrate our data. In particular, we use this result 
to determine the coefficient $e_2$
using our previous estimate for the velocity, $v_{s}$, finding $e_2=0.0841$. The remaining $\xi_K$ are then trivial to obtain.
In table~\ref{tab:xiKEgs} we list $D(J'_K)$ along with the resulting $\xi_K$. These independent (and absolute) estimates
of $\xi_K$ very nicely agree with the ones previously obtained in table~\ref{tab:NNN} clearly establishing that
the same length scale $\xi_K$ can be extracted from the scaling of the ground-state energy and the singlet-triplet
gap as one would expect from Kondo physics.

\begin{table}
  \centering
  \begin{tabular}{|c||cccccccc|}
    \hline\hline
      $J'_K$ &  1.000 &0.800 &0.600 &0.525 &0.450 &0.410 &0.370 &0.300 \\
\hline
      $D$ & 0.24093  &  0.54043  & 1.4853  &  2.5426  &  4.76    &  7.14   &   11.5  &    $\sim$ 35 \\
\hline
      $\xi_K$&$\equiv$0.65& 1.89& 5.81& 10.19 &19.4 &29.3& 47& $\sim$145 \\ 
  \hline\hline
  \end{tabular}
  \caption{The numerically determined values for $D$ for the SCKM at $J_2^c$ and the resulting $\xi_K$. \label{tab:xiKEgs}}
\end{table}

%
\section{Kondo effect in the nearest neighbor SCKM\label{sec:SCKM}, $J_2=0$}
We now turn to a discussion of the nearest neighbor SCKM, Eq.~(\ref{NN}), in the absence
of any second neighbor coupling ($J_2=0$). 
As outlined, the bulk marginal coupling in the SCKM, $g$, is in
this case non-zero, $g>0$.  
An exact Bethe ansatz solution of this model was developed by FZ~\cite{Frahm97}. 
We start by showing QMC results for this non-frustrated model, before presenting BA results for the singlet triplet gap. Then we discuss the connections with the usual Kondo physics.

\subsection{Impurity susceptibility}
\label{sec:chiimp}
We now want to compute the spin susceptibility of the SCKM defined 
in Eq.~(\ref{NN}). We take advantage of the non frustrated nature of this antiferromagnetic spin chain model to perform large scale 
QMC simulations using the Stochastic Series Expansion of the partition function in a loop algorithm framework (see Ref.~\cite{Sandvik02} for details). 
\subsubsection{Quantum Monte Carlo results}
\begin{figure}[!ht]
\begin{center}
\includegraphics[width=10cm,clip]{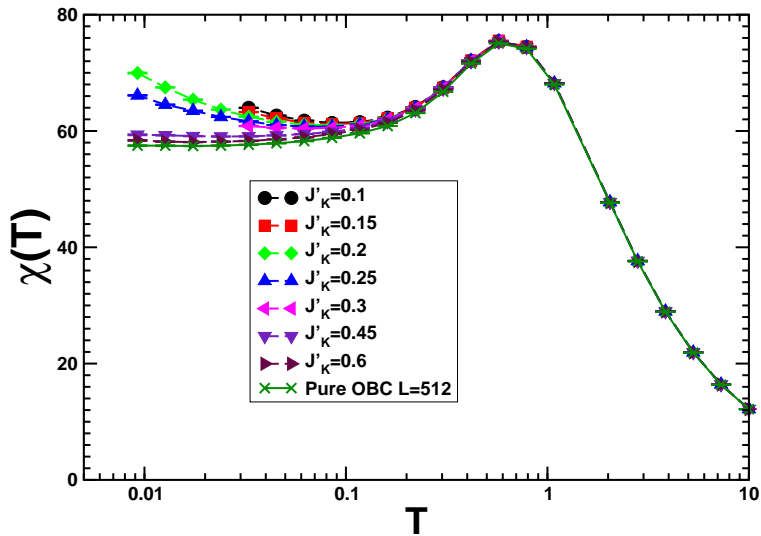}
\end{center}
\caption{Total susceptibility $\chi(T)$ obtained by Quantum Monte Carlo
simulations of the SCKM ($J_2=0$) with chains of length $L=512$ coupled to an
impurity spin with various Kondo couplings $J_K'$ indicated as various
symbols on the plot. Results for a pure open chain are also shown (solid
curve).}

\label{fig:ChiTotal}
\end{figure}

While the QMC simulations are done on finite size systems, on can still define the impurity susceptibility by 
\be
\chi_{\rm{imp}}=\lim_{L\to\infty}\left[\chi(L+1,J_K')-\chi(L,J_K'=1)\right],
\label{chiimpdef}
\ee
where $L+1$ corresponds to a system of $L$ spins in the bulk coupled to 1
boundary impurity with $J_K'$, and $L$ corresponds to a pure chain of $L$ spins
(see Fig. \ref{fig:NNchain}).  The QMC simulations have been carried out for $L=512$ at low
temperatures for various Kondo exchange couplings, $J_K'$.
While we are not in the
thermodynamic limit $L\to\infty$, we still expect for such a rather big system
size that one can capture all essential features of the Kondo physics present in
the spin chain model. 
Results for $\chi(513,J'_{K})$ are displayed in
Fig.~\ref{fig:ChiTotal} where a clear upturn is visible at low temperature.

As in the usual Kondo problem, one expects the impurity spin to be essentially
free at very high temperatures $T\gg T_{K}$, leading to a Curie-like divergence
$\sim 1/T$.  On the other hand, when the temperature is decreased, the
effective Kondo coupling starts to grow and the impurity eventually becomes
screened when $T\ll T_{K}$, forming a strongly entangled spin singlet with the
bulk, and thus resulting in an absence of features in the total susceptibility.

\subsubsection{Scaling properties and Kondo temperature}
\begin{figure}[!ht]
\begin{center}
\includegraphics[width=14cm,clip]{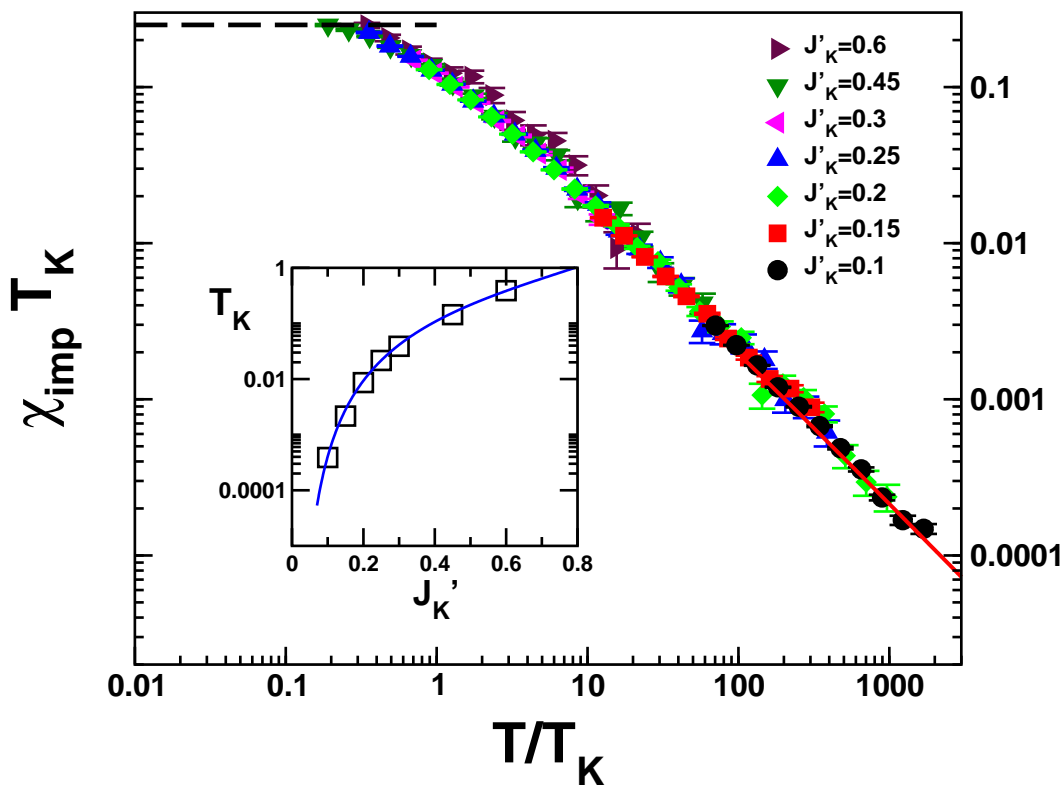}
\end{center}
\caption{Impurity susceptibility $\chi_{\rm imp}$ obtained by Quantum Monte Carlo calculations
of the SCKM ($J_2=0$) with chains of length $L=512$
coupled to an impurity spin with various Kondo couplings $J_K'$ indicated on the plot. 
The collapse for $\chi_{\rm imp}\times T_{K}$ was obtained starting from the weak coupling regime with the curve 
$\chi_{\rm imp}(T\gg T_{K})=0.25/T(1-1/\ln(T/T_K)$ (red line on the right) serving as a support for the 
rest of the collapse. The strong coupling limit $\chi_{\rm imp}(T\ll T_{K})=0.25/T_{K}$ is shown by the black horizontal dashed line on the right. Inset: Kondo temperature estimates obtained 
from the collapse on the main panel plotted versus the coupling $J_{K}'$. The solid curve 
is the expression Eq.~(\ref{eq:FZ}) with a prefactor $\simeq 5$.}
\label{fig:ChiTL512}
\end{figure}

As in the FEKM, one might expect the impurity susceptibility to be a scaling function of $T/T_{K}$
up to logarithmic corrections coming from the bulk marginal operator $g$.
More precisely, weak and strong coupling regimes are characterized by the following behaviors:
\be
4T_{K}\times \chi_{\rm imp}\to\left\{
\begin{array}{rl}
{T_{K}}/{T} & {\rm if}\  T\gg T_K\\
1 & {\rm{if}}\ T\ll T_K.\
\end{array}
\right.,
\ee
In the weak coupling regime, the lowest order perturbative expansion
\be
\chi_{\rm imp}=\frac{1}{4T}(1-1/\ln(T/T_K)+\ldots),
\ee
has been used to start building the scaling plot for $T_{K}\times \chi_{\rm imp}$,
as shown in Fig.~\ref{fig:ChiTL512} where the data of
Fig.~\ref{fig:ChiTotal} have been used to compute $\chi_{\rm imp}$. Step by
step, the temperature axis has been rescaled with the Kondo temperature
$T_{K}$ like $T\to T/T_{K}$ in order to produce the best collapse of the data
onto a curve which reproduces the expected behavior for the usual Kondo
problem (see Ref.~\cite{Hewson} for instance). 
Ignoring scaling violations arising from the bulk marginal
coupling (which we expect to be small), one expects the scaling plot shown in
Fig.~\ref{fig:ChiTL512} to give a rather good estimate of the Kondo
temperature $T_{K}$ for the Heisenberg spin chain.  The resulting estimates for $T_K$
(and thus $\xi_K=v_s/T_K$ with $v_s=\pi/2$ for the Heisenberg chain) are
listed in the table~\ref{tab:NN}, and plotted in the inset of
Fig.~\ref{fig:ChiTL512}. One sees the Kondo energy scale $T_{K}$ going to
zero exponentially when $J'_{K}\to 0$. We have successfully compared these
estimates to the FZ form~\cite{Frahm97}

\be
\label{eq:FZ}
T_{K}\propto\exp(-\pi{\sqrt{1/J'_{K}-1}}),
\ee
which in the limit of small Kondo coupling $J'_{K}$ is equivalent to the exponential square-root form Eq.~(\ref{xiKapprox}) derived in the previous section from the RG.

One technical remark has to be mentioned about the QMC data for $\chi_{\rm imp}$. 
At the strong coupling fixed point it is quite difficult to simulate the system 
at very low temperature $T\ll T_{K}$ {\it and} in the scaling limit $T\gg v_{s}/L$ at the same time.
Finally, as briefly mentioned above, we expect that the bulk marginal interaction will lead to small corrections to the 
susceptibility.
\begin{table}
\begin{center}
\begin{tabular}{|r||r|r|r|r|r|r|r|}
\hline
$J_K'$&0.1&0.15& 0.2&0.25&0.3&0.45&0.6\\
\hline
$T_K\sim$&0.00039&0.0086&0.022&0.039&0.0144&0.1&0.388\\
\hline
$\xi_K\sim$&4040&727&182&73&40&11&4\\
\hline
\end{tabular}
\caption{Kondo temperature $T_K$ and the associated Kondo length scale $\xi_K=\pi/(2T_K)$ for the SCKM ($J_2=0$), Eq.~(\ref{NN}), with a Kondo coupling $J_K'$.  The values are estimated from the impurity susceptibility data (see Fig.~\ref{fig:ChiTL512}).}
\label{tab:NN}
\end{center}
\end{table}
\subsection{Bethe ansatz solution}\label{sec:BAE}
We start from an open Heisenberg chain with a boundary spin impurity Eqs.~(\ref{NN}) and an external magnetic field 
${\vec{B}}=B{\vec{e_z}}$ 
\be
\H=\sum_{i=1}^{L-1}{\vec{S}}_{i}\cdot{\vec{S}}_{i+1}+J_K' {\vec{S}}_{imp}\cdot{\vec{S}}_1-\sum_{i=1}^{L}BS_{i}^{z}.
\ee
This model has been shown to be integrable~\cite{Frahm97} and allows a 
Bethe Ansatz solution which, in the case where the impurity spin 
$S_{imp}=1/2$, reads 
\bea
[e_1(\Lambda_j)]^{2L}
e_1(\Lambda_j+{\tilde{\lambda}}_K)
e_1(\Lambda_j-{\tilde{\lambda}}_K)
= \prod_{k=1,k\ne j}^{M}e_2(\Lambda_j-\Lambda_k)e_2(\Lambda_j+\Lambda_k).\nonumber\\
\label{eq:BA2}
\eea
$M$ is the number of down spins, the function $e_{n}(x)=\frac{x+ni/2}{x-ni/2}$ and 
the impurity coupling constant $J_K'$ is related to 
${\tilde{\lambda}}_K$ by
\be
J_K'=\frac{1}{1+{\tilde{\lambda}}_{K}^{2}}.
\ee
Note that the $\tilde \lambda_K$ defined here {\it is not} the same as 
the $\lambda_K$ defined in Sec. II. 
It is more convenient to re-write Eq.~(\ref{eq:BA2}) as
\bea
\left[\frac{\Lambda_j+\frac{i}{2}}{\Lambda_j-\frac{i}{2}}\right]^{2L}
\left[\frac{\Lambda_j+{\tilde{\lambda}}_K+\frac{i}{2}}{\Lambda_j+
{\tilde{\lambda}}_K-\frac{i}{2}}\right]
&\left[\frac{\Lambda_j-{\tilde{\lambda}}_K+\frac{i}{2}}
{\Lambda_j-{\tilde{\lambda}}_K-\frac{i}{2}}\right]=\nonumber\\
&\prod_{k=1,k\ne j}^{M}
\left[\frac{\Lambda_j+\Lambda_k+i}{\Lambda_j+\Lambda_k-i}\right]
\left[\frac{\Lambda_j-\Lambda_k+i}{\Lambda_j-\Lambda_k-i}\right].
\label{eq:BA1}
\eea
Then we get
\bea
2L\theta(2\Lambda_j)+\theta(2\Lambda_j+2{\tilde{\lambda}}_K)
&+\theta(2\Lambda_j-2{\tilde{\lambda}}_K)
=\nonumber\\
&\sum_{k=1,k\ne j}^{M}\theta(\Lambda_j+\Lambda_k)
+\theta(\Lambda_j-\Lambda_k)
-\pi I_{j},
\eea
where $\theta(x)=\arctan\left({x}\right)$, 
the $I_j$ are all integers and $j=1,...,M$. 
The total energy in an external magnetic field $B$ can be expressed~\cite{Frahm97}
\be
E(M)=\frac{L-1}{4}+\frac{J_K'}{4}-\frac{1}{2}\sum_{j=1}^{M}\frac{1}{\Lambda_{j}^{2}+1/4}-\left(\frac{L+1}{2}-M\right)B.
\ee
\subsection{Singlet-triplet gap}

\begin{figure}[b]
\begin{center}
\includegraphics[width=10cm,clip]{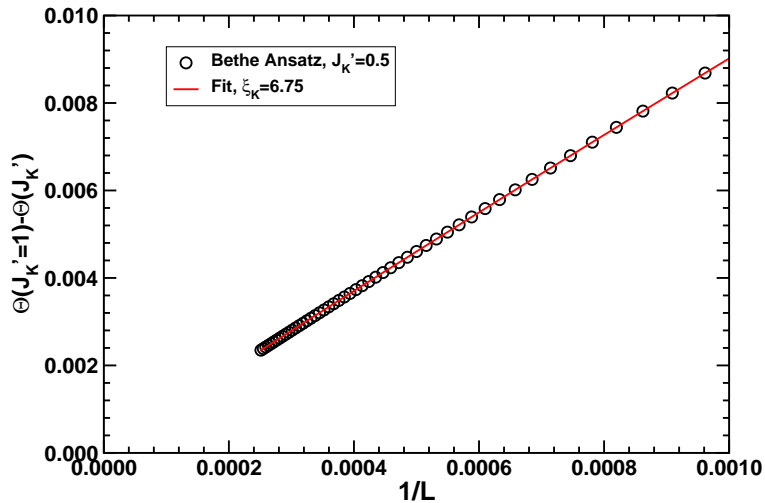}
\end{center}
\caption{Bethe Ansatz results for the SCKM ($J_2=0$). Data are shown for the difference $\Theta(J_K'=1)-\Theta(J_K')$ [Eq.~(\ref{eq:thetaprime})]
computed exactly for system sizes $20\le L\le 4000$ with an impurity coupling $J_K'=0.5$. The solid line represents a fit to the data
of the form Eq.~(\ref{eq:thetaprime}) with $\xi_K=6.75, a=-10.21, b=-109.39$.}
\label{fig:DDelta}
\end{figure}
%
For the $J_2=0$ spin chain with open boundary conditions, the result Eq. (\ref{eq:THETA0}) 
is modified, to first order in the marginally irrelevant bulk coupling constant, $g(L)$ to \cite{AQ}:
\be
\Delta_{ST}(L)=\frac{\pi v_s}{L}[1- g(L)]
\label{eq:Delta}
\ee
We first compute $\Delta_{ST}(L)$ in the clean case ($J'_{K}=1$) using the Bethe Ansatz solution up to $L=10^{4}$ 
to extract the marginal coupling $g(L)$ from Eq.~(\ref{eq:Delta}). 
A very good agreement (for $L\gg 100$) with 
$g(L)$ from Eq.~(\ref{g(L)2}) is observed using $L_{1}\simeq 0.92$.

For $J_K\ne 1$ we expect Eq.~(\ref{eq:Delta}) to continue to hold in the strong coupling limit of the SCKM,
$\xi_K\ll L$, but we may also include the Fermi liquid correction of Eq. (\ref{eq:THETA1}):
\be
\Delta_{ST}(L\gg\xi_K)=\frac{\pi v_s}{L} \left[ 1-g(L)-\frac{\pi\xi_K}{2L}\right] .
\label{eq:FLTH}
\ee
Note that this result is to first order in $g(L)$ and first order in $\xi_K/L$ only. 
It should be corrected by a Taylor series in these 2 coupling constants.  We also emphasize that the
$g(L)\sim 1/\ln(L)$ term is the dominant correction in this expression. Hence, in order to verify the
presence of the term proportional to $\xi_K$ it is natural to study the following quantity:
\bea
\frac{L}{\pi v_s}\Delta'_{ST}(J_K')&=&
\frac{L}{\pi v_s} \left[ \Delta_{ST}(J_K'=1)-\Delta_{ST}(J_K') \right]\nonumber\\
  &\equiv &\Theta(J'_K=1)-\Theta(J'_K) \nonumber\\
&\sim&\frac{\pi(\xi_K-\xi_K(J_K'=1))}{2L}+\frac{a}{L\ln(L)}+\frac{b}{L^2},  \ \ L\gg \xi_K\label{eq:thetaprime}
\eea
where we have included the leading terms in the expected Taylor series in the 2 coupling constants.
In particular, one might expect a term of the form $\sim (\pi v_s/L)g(L) \xi_K/L~\sim 1/(L^2\ln(L))$ to be present in 
the finite-size corrections to $\Delta_{ST}(L)$. The analytical calculation of the coefficient of such a term
would be challenging since it would involve diagrams with 6 current operators.
In the opposite limit $L\ll \xi_K$
we can calculate $\Delta_{ST}(L)$ in perturbation theory in $\lambda_K$, Eq. (\ref{eq:THETA2}).  
At weak coupling ($L\ll \xi_K$) there is no correction of $O(g)$. 
This follows because both singlet and triplet states 
arise from the same state (with $S=1/2$) of the chain of $L$ spins.

\begin{figure}[b]
\begin{center}
\includegraphics[width=10cm,clip]{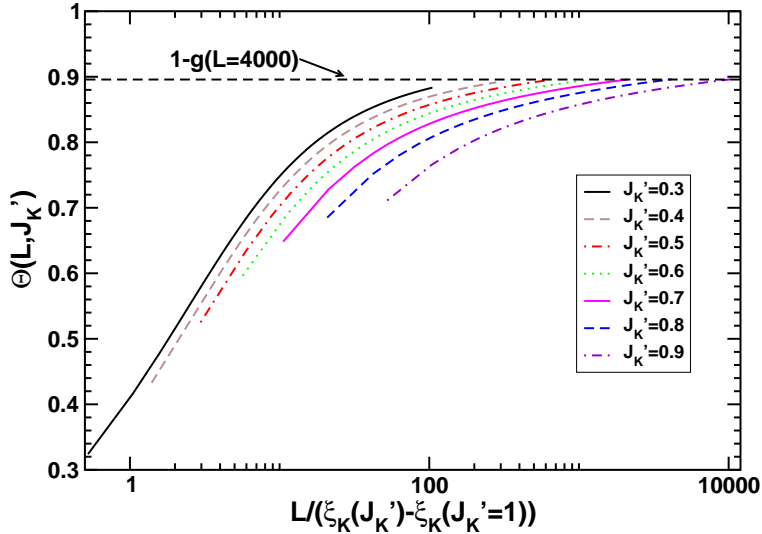}
\end{center}
\caption{Bethe Ansatz results for the SCKM ($J_2=0$). Data are shown for
the singlet-triplet gap $\Delta_{ST}$ versus $L/(\xi_K-\xi_K(J_K'=1))$
computed exactly for various system sizes $20\le L\le 4000$ and impurity couplings $J_K'=0.3,\ldots,0.9$ as indicated in the plot 
by the different symbols. $\xi_K-\xi_K(J_K=1)$ is obtained from 
fitting to Eq.~(\ref{eq:thetaprime}) as listed in table~\ref{tab:NNBA}.
Due the the presence of the marginal coupling, $g$, the data clearly do not scale with $L/\xi_K$ as was the
case at the critical point $J_2^c$ as shown in Fig.~\ref{fig:THETANNN}.
}
\label{fig:ThetaNN}
\end{figure}
Exact results available form the Bethe ansatz solution of finite systems allow us to verify the
presence of the $\pi\xi_K/(2L)$ term in the singlet-triplet gap, Eqs.~(\ref{eq:FLTH}), (\ref{eq:thetaprime}).
Our results for $\Theta(J'_K=1)-\Theta(J'_K)$ are shown in Fig.~\ref{fig:DDelta} for $J_K'=0.5$, clearly demonstrating
the presence of a linear term $\sim (\xi_K-\xi_K(J_K'=1))/L$. 
If the higher order corrections described by Eq.~(\ref{eq:thetaprime}) are
included a very precise absolute estimate for $\xi_K-\xi_K(J_K'=1)$ can be obtained. 
We expect $\xi_K(J_K'=1)$ to be ${\cal O}(1)$ and therefore to be negligible compared to $\xi_K$ for large
enough $\xi_K$.
Fits to the form given by Eq.~(\ref{eq:thetaprime})
are extremely good down to $J_K'=0.3$ and the obtained values of $\xi_K-\xi_K(J_K'=1)$ 
are listed in table~\ref{tab:NNBA}. For smaller
values of $J_K'$ the diverging $\xi_K$ prohibits reliable fits for the system sizes where we can numerically solve
the Bethe ansatz equations.

\begin{table}
\centering
\begin{tabular}{|r||ccccccc|}
\hline\hline
$J_K'$&
  $0.3$   &
  $0.4$   &
  $0.5$   &
  $0.6$   &
  $0.7$   &
  $0.8 $  &
  $0.9$   \\
\hline
$\xi_K-\xi_K(J_K'=1)$
&$ 38.12$
&$ 14.32$
&$ 6.75$
&$ 3.53$
&$ 1.90$
&$ 0.96$
&$ 0.38$\\
\hline\hline
\end{tabular}
\caption{Bethe Ansatz estimates for the Kondo length scale $\xi_K-\xi_K(J_K'=1)$ of 
the SCKM Hamiltonian Eq.~(\ref{NN}), $J_2=0$, with a Kondo coupling $J_K'$.}
\label{tab:NNBA}
\end{table}

As we showed in Fig.~\ref{fig:THETANNN}, the quantity
$\Theta=L /(\pi v_s)\Delta_{ST}(L,J_K')$ scales with $L/\xi_K$ in the FEKM and also for the SCKM at $J_2^c$
where the bulk marginal coupling, $g$, is zero. At $J_2=0$, $g$ is non-zero and is in fact the dominant correction
term in singlet-triplet gap, Eq.~(\ref{eq:FLTH}). Hence, for $J_2=0$ we no longer expect scaling of $\Theta$ with
$L/\xi_K$, but instead, $\Theta$ should approach $1-g(L)$ for large $L$. This is demonstrated in Fig.~\ref{fig:ThetaNN}
where exact Bethe ansatz results for $\Theta$ for the SCKM at $J_2=0$ are plotted versus $L/(\xi_K-\xi_K(J_K'=1)$ with
$\xi_K-\xi_K(J_K'=1)$ obtained
from fitting to Eq.~(\ref{eq:thetaprime}) as listed in table~\ref{tab:NNBA}. Clearly the data do not scale, even for
sizable values of $\xi_K-\xi_K(J_K'=1)$ where $\xi_K(J_K'=1)$ can be neglected, but approach
the value $1-g(L)$ plotted as the dashed line with $g(L)$ from Eq.~(\ref{g(L)2}) using the previously determined 
$L_1\simeq0.92$. Data are shown for $20\le L\le 4000$.
From Eq.~(\ref{eq:FLTH}) it would be tempting to assume that the quantity $\Theta(J'_K=1)-\Theta(J'_K)$ 
should scale with $L/\xi_K$ for the SCKM at $J_2=0$, however, higher order terms in $g(L)$ not included in Eq.~(\ref{eq:FLTH})
do not cancel, corrupting the scaling. Likewise, we also do not expect $\Theta$ to scale with $L/\xi_K$ in the weak coupling
regime $L\ll \xi_K$ since, even though there is no correction of ${\cal O}(g)$, since $\lambda_K(L)$, Eq.~(\ref{xiKbeta1}),
is not a pure scaling function of $\xi_K/L$, unlike in the FEKM.

\subsection{Finite size scaling of the ground state energy in the SCKM with $J_2=0$}\label{sec:EgsJ20}
In section~\ref{sec:EgsJ20}, Eq.~(\ref{eq:Egs}),
we derived the finite size corrections to the ground-state energy
neglecting the bulk marginal operator, $g$. We now discuss the more general case 
where $g>0$ and logarithmic corrections arising from this operator are present.
For the uniform ($J_K'=1)$ chain with open boundary conditions these corrections are known~\cite{AQ}. 
In addition, as outlined
in \ref{app:gxik}, it is also possible to derive the universal term in the ground-state energy arising
from the coupling between $g$ and $\xi_K$:
\begin{equation}
\delta E_{GS}={\xi_Kgv_{s}\pi^2\over 32L^2}.
\label{eq:degsgxik}
\end{equation}
Combining the various results, we find for general $J_2\leq J_2^c$ and $J_K'\neq 1$ with $L\gg \xi_K$:
\begin{eqnarray}
E_{\rm GS}(L,J_K^\prime)=e_{0}(J_K')&+&e_{1}L-\frac{\pi v_{s}}{24L}\left[1-\frac{9g^2}{2}\right]\nonumber\\
&+& \left( e_{2}+\frac{\pi ^{2}v_{s}\xi _{K}}{48}\left(1+(3/2)g\right)\right) \frac{1}{L^{2}}.
\label{eq:TotalEgs}
\end{eqnarray}
Here, as $L\to\infty$, $g\sim 1/\ln L$.

We have attempted to verify the correction to the ground-state energy,
Eq.~(\ref{eq:degsgxik}), for non-zero $g$ through the use of the Bethe ansatz
solution for the SCKM at $J_2=0$~\cite{Frahm97}. Although a logarithmic term of
the correct order can clearly be identified, the sign appears to be the
opposite of Eq.~(\ref{eq:degsgxik}). This may be due to significant higher order corrections that could change the effective sign observed in the numerical results.

\subsection{Bethe ansatz results for the  magnetization, $M(H)$}\label{sec:Mimp}
FZ gave results on the  magnetization for the SCKM at zero temperature 
at infinite length using the Bethe Ansatz. 
They found \cite{Frahm97,Frahm07} that, at large $L$, the magnetization had the form:
\be M(L)=m_0L+(M^z_{edge}+M^z_{imp}) + O(1/L).\ee
Here $m_0$ is the bulk magnetization per unit length, which is linear in $H$ (up to 
logarithmic corrections) at $H\ll J$. For $H\ll J$, FZ found:
\be M^z_{edge}={1\over 4}\left[{1\over \ln (H_0/H)}-{1\over 2}{\ln [(1/2)\ln (H_0/H)]\over \ln^2 (H_0/H)}
\right]+\ldots .\ee
FZ wrote $M^z_{imp}$ in terms of a field
\be H_-\equiv H_0e^{- \pi |\tilde{\lambda}_K|}.\ee
Here:
\be H_0\equiv J\sqrt{\pi^3/e}\ee
and $\tilde{\lambda}_K$ is defined by:
\be J_K'={1\over 1+\tilde{\lambda}^2_K}.\ee
They identified $H_-$ with a Kondo temperature, $T_K$. 
At $J_K'\ll 1$ this has the exponential square-root form derived 
in section~\ref{sec:FT} from the RG:
\be T_K=H_0e^{-\pi \sqrt{1/J_K'-1}}.\label{eq:TKFZ}\ee
 At $H\ll T_K$ FZ
found that $M^z_{imp}$ vanishes (for an S=1/2 impurity). 
However, for $T_K\ll H\ll J$, they found:
\bea M^z_{\rm{imp}}(H)&\to& 
{1\over 2}
 - {1\over 4}\left[{1\over \ln (H/T_K)}+{\ln [(1/2)\ln (H/T_K)]\over \ln^2 (H/T_K)}
\right]\nonumber \\
&-&{1\over 4}\left[{1\over \ln (H_0^2/HT_K)}+{1\over 2}{\ln [(1/2)\ln (H_0^2/HT_K)]\over \ln^2(H_0^2/HT_K)} 
\right] ,\nonumber \\
&& (T_K\ll H\ll J).\label{mzimp}\eea

For the model with $J_K'=1$, Furusaki and Hikihara \cite{Furusaki} 
observed that 
$M^z_{edge}$  (which was obtained earlier by Bethe ansatz for the $J_K'=1$ case in related models 
\cite{Essler,Asakawa}) 
can be derived from renormalization group improved first order perturbation theory in $g$.
It is simply $(1/4)g(H)$, up to a higher order term (at small $H$) of $O[1/\ln^2 (H_0/H)]$. 
We should associate terms $(1/8)g(H)$ with each boundary. 

It is interesting to consider the ``impurity magnetization'' ${\cal M}_{imp}$ as 
it is usually defined in the theory of the Kondo effect.  This 
is the  additional magnetization resulting from adding the impurity.   For the 
spin chain this is naturally defined as:
\be {\cal M}_{imp}\equiv \lim_{L\to \infty}[M(J_K,L+1)-M(J_K=1,L)].\ee
Ignoring a term $m_0(H)$, the bulk magnetization per unit length, 
which is smaller by a factor of $(H/J)$ than the other contributions, 
we see that:
\be {\cal M}_{imp}=M^z_{imp}.\ee
The first line in ${\cal M}_{imp}$ in Eq. (\ref{mzimp}) is precisely the 
standard result for the FEKM, at $T_K\ll H\ll D$, up to 
corrections of $O[1/\ln^2(H/T_K)]$. (See, for example, \cite{Hewson}.) It is, of course, a scaling function 
of $H/T_K$. The second line in ${\cal M}_{imp}$ in Eq. (\ref{mzimp}) 
is rather complicated. 
 Using:
\be \ln (HT_K/H_0^2)=2\ln (H/H_0)+\ln (T_K/H)\ee
we can write it in terms of $g(H)$ and $\lambda_K(H)$. 
\be \delta M_{imp} \approx {-1/4\over 2/g(H)+1/\lambda_K(H)}.\label{nonan}\ee
This could be Taylor expanded in powers of $g(H)$ and indeed has 
the form of Eq. (\ref{scalcorr}). However, we find 
the peculiar non-analytic form of Eq. (\ref{nonan}) puzzling. 
\be {1\over 2/g+1/\lambda_K}={g\over 2}\sum_{n=0}^\infty \left({-g\over 2\lambda_K}\right)^n.\ee
Clearly, this expression does not have a joint Taylor expansion in $g$ and $\lambda_K$.

\section{Conclusions}\label{sec:conclusion}

\begin{figure}[t]
\begin{center}
\includegraphics[width=14cm,clip]{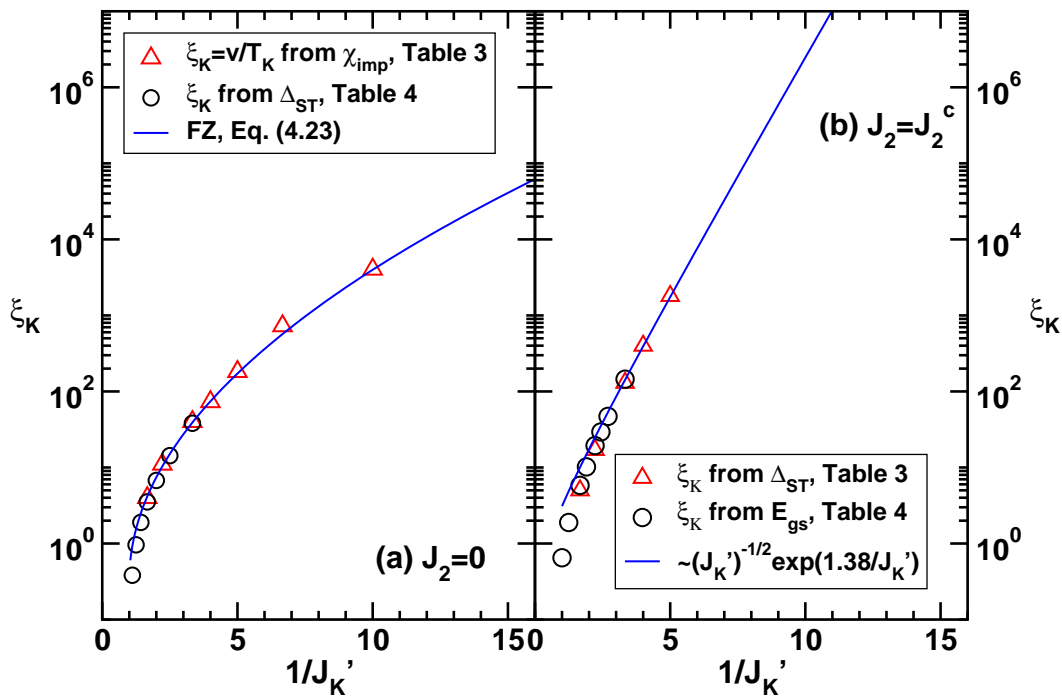}
\end{center}
\caption{(a) Results for the nearest neighbor SCKM, $J_2=0$. $\xi_K=v_{s}/T_K$ as obtained from the impurity susceptibility, $\chi_{imp}$ (table~\ref{tab:NN}) shown
with $\xi_K$ obtained from scaling of $\Delta_{ST}$ (table~\ref{tab:NNBA}). 
Both quantities are seen to agree with $v/T_K$ from
FZ, Eq.~(\ref{eq:TKFZ}). (b) Results for the next-nearest neighbor SCKM with $J_2=J_2^c$. $\xi_K$ are shown as obtained from
scaling of $\Delta_{ST}$ and from the ground-state energy. Agreement is observed with the RG result, Eq.~(\ref{eq:XiKNNN}).}
\label{fig:AllXiK}
\end{figure}

Our main results for $\xi_K=v_{s}/T_K$ are summarized in Fig.~\ref{fig:AllXiK} for the SCKM at
$J_2=0$, panel (a), and $J_2=J_2^c$, panel (b). 
We have shown that the spin chain Kondo model with the second 
neighbor coupling, $J_2$ adjusted to the critical point has 
the same low energy, long distance behavior as the usual 
free electron Kondo model. This leads to the usual exponential divergence of $\xi_K$ with $1/J_K'$
as shown in panel (b) where the concordance of $\xi_K$ as obtained from scaling of $\Delta_{ST}$, $E_{gs}$
and RG is demonstrated. Furthermore, at $J_2=J_2^c$ many quantities display scaling with $L/\xi_K$ in an
identical manner for the SCKM and FEKM (Fig.~\ref{fig:THETANNN}).  For $J_2<J_2^c$
a new kind of Kondo effect arises in the SCKM due to the presence of the non-zero marginal coupling $g$. As shown
in panel (a) for $J_2=0$ this leads to a more slowly diverging $\xi_K\sim \exp(\sqrt{1/J_K'-1})$ in agreement with
the analytical Bethe ansatz result~\cite{Frahm97}. The non-zero marginal coupling $g>0$ destroys scaling with
$\xi_K/L$ and reliable estimates for $\xi_K$ can only be obtained from a numerical determination of finite-size
corrections. However, the strong coupling fixed point remains that of a completely screened impurity and the usual
Kondo physics is still present in this case albeit obscured by significant logarithmic corrections arising from
the bulk marginal operator, $g$.
Hence, we expect physical quantities to behave the same way as for 
the FEKM with corrections which only vanish logarithmically 
with decreasing energy or increasing system size:
\be f(L,J_K')=f_0(L/\xi_K)+g(L)f_1(L/\xi_K)+g^2(L)f_2(L/\xi_K)+\ldots, \ee
where $f_0(L/\xi_K)$ is the universal scaling function 
occurring for the FEKM. 

Several open problems pertaining to the Kondo effect as it occurs in the SCKM at $J_2=0$ ($g>0$) remain.
First, as briefly mentioned in section~\ref{sec:chiimp} we expect that the bulk marginal interaction will give rise to
small corrections to the susceptibility, precise analytical or numerical results for such corrections would be desirable.
Secondly, in section~\ref{sec:EgsJ20} we derived an analytical result for the coefficient of the $(\pi v_s/L)g(L)\xi_K/L$ term
in the ground-state energy, Eq.~(\ref{eq:degsgxik}), lending support to the proposed form
of the scaling corrections Eq.~(\ref{scalcorr}).
However, even though our numerical solution of the Bethe ansatz
equations show a logarithmic term of the correct order the sign appears to be the opposite of that of Eq.~(\ref{eq:degsgxik}).
A possible explanation for this would be that higher order are significant and change the effective sign seen in the
numerics. Further analytical insight to the higher order terms or drastically improved numerical results would be valuable
to resolve this questions. 
Thirdly, additional support for the proposed scaling corrections, Eq.~(\ref{scalcorr}), in the form of an analytical calculation
of the coefficient of the $(\pi v/L)g(L)\xi_K/L$ term in $\Delta_{ST}$ would also be of considerable interest. We have
so far not been able to derive this coefficient in a simple manner. 
As pointed out in section~\ref{sec:Mimp}, we find the peculiar non-analytic form of Eq. (\ref{nonan}) puzzling since
it does not have a joint Taylor expansion in $g$ and $\lambda_K$, an issue we hope will be resolved in the near future.
Finally, we hope that the spin chain Kondo models studied in this paper will help experimentalists for a better understanding of impurity effects in spin chains materials.

\ack
We are grateful to M.-S. Chang, R. Pereira for interesting discussions and H. Frahm, A. Furusaki
for helpful correspondence. This research was
supported by NSERC (all authors), the Swiss National Fund (NL), MaNEP (NL), CIfAR (IA), CFI (ESS), and SHARCNET
(ESS). ESS gratefully acknowledge the hospitality
of the Kavli Institute for Theoretical Physics in Santa
Barbara, where part of this work was carried out and supported by
the NSF under Grant No. PHY05-51164.
Numerical simulations have been performed on the WestGrid network
and the SCHARCNET facility at McMaster University.

\appendix

\section{Derivation of $g(L)\xi_K/L^2$ term in ground state energy}\label{app:gxik}
We follow closely \cite{AQ}, which we refer to as AQ, but note the change in notation:
\be g_{AQ}=(\sqrt{3}/4\pi)g,\  \   \vec J_{AQ}=(1/2\pi )\vec J.\ee
First let us correct (2.21) of AQ.  This was meant to be the 
connected part but the subscript ``connected'', as in (2.18) was missing. 
More substantially, the connected Green's function should have 
two other terms, missing from (2.21).  The correct result, with $\vec J$ 
normalized as we do is:
\bea 
\langle(\vec J(z_1)\cdot \vec J(z_2))J^a(0)J^b(\tau )\rangle_C&=&
{\delta^{ab}\over z_1z_2(\tau -z_1)(\tau -z_2)}
\nonumber\\
&+&{\delta^{ab}\over 4z_1^2(\tau -z_2)^2}+{\delta^{ab}\over 4z_2^2(\tau -z_1)^2}.
\label{corr}\eea
(We dropped the $L$ subscripts but all currents are left-moving.)
These last 2 terms, missing from (2.21) of AQ, do not contribute to the $g^2/L$ term in $E_{GS}$ 
although they do lead to an additional non-universal $g^2$ term without the $1/L$ factor. 
Therefore, they make no change in the conclusions of AQ. 

Now consider including the Fermi liquid interaction as well as the 
bulk marginal interaction:
\be H=H_0-\frac{\xi_K}{6}\vec J^2(0)-\frac{g}{2\pi }\int_0^Ldx \vec J(x)\cdot \vec J(-x).\label{ham}\ee
For infinite $L$, standard perturbation theory to first order in $g$ and $\xi_K$ gives:
\be \delta E_{GS}=-\frac{\xi_K}{6}\frac{g}{2\pi }\int_{-\infty}^\infty d\tau \int_0^\infty dx 
\langle\vec J^2(0)\vec J(x,\tau )\cdot \vec J(-x,\tau )\rangle_C.\label{pert}\ee
From Eq. (\ref{corr}) setting $\tau \to 0$, $z_1\to \tau +ix$, $z_2\to \tau -ix$, 
and summing over $a=b$, this is:
\be 
\delta E_{GS}=-\frac{\xi_K}{6}\frac{g}{2\pi }\int_{-\infty}^\infty d\tau \int_0^\infty dx 
{9\over 2(\tau+ix)^2(\tau -ix)^2}.\ee
Note that all 3 terms in Eq. (\ref{corr}) contribute with 
the 2 last terms giving 1/2 the contribution of the first term.
Also note that the integrand is an even function of $x$ so we can extend the 
region of integration to $-\infty <x<\infty$. 
 Doing the $x$-integral 
gives:
\be \delta E_{GS}=-\frac{\xi_K}{6}\frac{g}{2\pi }\frac{9\pi}{ 8}\int_{-\infty}^\infty {d\tau \over |\tau |^3}.\ee
This integral is divergent.  We cut it off, $|\tau|>\tau_0$, giving:
\be 
\delta E_{GS}=-{3\xi_Kg\over 32\tau_0^2}.\label{nonun}\ee
This is a non-universal term with no  $L$-dependence.  To get 
the $L$-dependent term we need the generalization of Eq. (\ref{corr}) to 
finite $L$.  This is obtained by:
\be \tau\pm ix \to (2L/\pi )\sinh [(\tau \pm ix)\pi /2L].\ee
the $x$ and $\tau$ integrals (with the cut off on the $\tau$ integral) can again be done exactly.
Again the integrand is an even function of $x$ so we can extend it to $-L<x<L$ and 
then change variables to $z=e^{i\pi x/L}$.  The resulting contour integral is elementary, 
with a double pole inside the contour
 giving:
\be \delta E_{GS}=-{3\xi_Kg\pi^3\over 32L^3}\int_{-\infty}^\infty d\tau {\cosh (\pi \tau /L)
\over |\sinh (\pi \tau /L)|^3}.\ee
Doing the cut-off $\tau$ integral exactly gives:
\be \delta E_{GS}=-{3\xi_Kg\over 32[(L/\pi )\sinh(\pi \tau_0/L)]^2}.\label{exact}\ee
Now we Taylor expand to second order in $\tau_0/L$ giving:
\be \delta E_{GS}=-{3\xi_Kg\over 32}\left[{1\over\tau_0^2}-{\pi^2\over 3L^2}+\ldots \right].\ee
So the desired universal term is:
\be \delta E_{GS}={\xi_Kg\pi^2\over 32L^2}.\ee
Reinserting the factor of $v_{s}$ which was set to 1 
gives the ${\cal O}(g/L^2)$ term in Eq.~(\ref{eq:TotalEgs}).

\section{Extracting the $1/L^2$ Term in $E_{GS}$}\label{app:egs}
In this appendix we describe a form of Richardson extrapolation that is
useful for isolating the $1/L^2$ term in the ground-state energy needed to
extract $\xi_K$ in Eq.~(\ref{eq:Egs}).

We assume that the ground-state energy is of the form, Eq.~(\ref{eq:Egs}):
\begin{equation}
E^{(0)}(\RR)=Ac_{-1}^0(\RR)+Bc_0^0(\RR)+Cc_1^0(\RR)+Dc_2^0(\RR)+\mathcal{O}(\RR^{-3}),
\end{equation}
with $c_{-1}^0(\RR)=\RR, c_0^0(\RR)=1, c_1^0(\RR)=1/\RR$ and $c_2^0(\RR)=1/\RR^2$.
We're interested in determining the coefficient $D$. Furthermore,
we assume that $E(\RR)$ is known for a sequence of $\RR$. It is then easiest
to proceed using a form Richardson extrapolation by systematically removing the $A,B,C$
terms. We eliminate the term proportional to $A$ by constructing a new series:
\begin{equation}
E^{(1)}(\RR)=E^{(0)}(\RR)-E^{(0)}(\RR+2)\frac{c_{-1}^0(\RR)}{c_{-1}^0(\RR+2)}
\end{equation}
Naturally, $E^{(1)}$ must be of the form:
\begin{equation}
E^{(1)}(\RR)=Bc_0^1(\RR)+Cc_1^1(\RR)+Dc_2^1(\RR)+\mathcal{O}(\RR^{-2}),
\end{equation}
with
\begin{equation}
c_i^1(\RR)=c_i^0(\RR)-c_i^0(\RR+2)\frac{c_{-1}^0(\RR)}{c_{-1}^0(\RR+2)},\ i=0,1,2.
\end{equation}
Now we can eliminate the term proportional to $B$ by transforming:
\begin{equation}
E^{(2)}(\RR)=E^{(1)}(\RR)-E^{(1)}(\RR+2)\frac{c_{0}^1(\RR)}{c_{0}^1(\RR+2)},
\end{equation}
which we can write as:
\begin{eqnarray}
E^{(2)}(\RR)&=&Cc_1^2(\RR)+Dc_2^2(\RR^{2})+\mathcal{O}(\RR^{-1})\nonumber\\ 
c_i^2(\RR)&=&c_i^1(\RR)-c_i^1(\RR+2)\frac{c_{0}^1(\RR)}{c_{0}^1(\RR+2)}, i=1,2.\nonumber\\
\end{eqnarray}
Repeating once more we obtain $E^{(3)}(\RR)$ of the form:
\begin{eqnarray}
E^{(3)}(\RR)&=&Dc_2^3(\RR^{2})+\mathcal{O}(\RR^{-1})\nonumber\\ 
c_2^3(\RR)&=&c_2^2(\RR)-c_2^2(\RR+2)\frac{c_{1}^2(\RR)}{c_{1}^2(\RR+2)}.\nonumber\\
\label{eq:E3}
\end{eqnarray}
From which the constant $D$ can be extracted by plotting $E^{(3)}(\RR)/c_2^3(\RR)$.

\newpage
\bibliographystyle{unsrt}
\bibliography{keac}
\end{document}